# Guiding without Generating: Artificial Intelligence (AI)-Enabled Topic Nudges in Online Reviews

Fangyan Wang*
Mitch Daniels School of Business, Purdue University
wang6123@purdue.edu

Sai Liang
College of Tourism and Service Management, Nankai University
liangsai@nankai.edu.cn

Zaiyan Wei
Mitch Daniels School of Business, Purdue University
zaiyan@purdue.edu

This Draft: April 2026

* For helpful comments, we are grateful to Daisy Dai, Alan Dennis, Sung Joo Kim, Alex Moehring, and Agrim Sachdeva. We also thank seminar attendees at Purdue University, the University of Arizona, and McGill University, as well as participants at and/or reviewers for 2024 KDSA Research Symposium, 2024 Conference on Artificial Intelligence, Machine Learning, and Business Analytics (AIML), the 5th AI in Management Conference, 2025 Biz AI Conference: AI Applications in Business Research, the 18th China Summer Workshop on Information Management (CSWIM), and 2025 Fisher AI in Business Conference. Zaiyan Wei would like to thank the funding support from Mitch Daniels School of Business. All results have been reviewed to ensure that no confidential information is disclosed. The authors contribute equally.

# Guiding without Generating: Artificial Intelligence (AI)-Enabled Topic Nudges in Online Reviews

**Abstract**

Digital platforms increasingly face a common challenge in the age of artificial intelligence (AI): how to elicit richer and more useful user-generated content (UGC) without fully automating content production. We study this question in the context of online reviews by examining Yelp's introduction of an AI-enabled topic nudging tool in 2023. The feature provides real-time prompts to guide review writers in addressing key dimensions of the dining experience as they compose their reviews. Using more than 1.5 million Yelp reviews and a differences-in-differences design, we find that AI-enabled topic nudges significantly reshape the generation of reviews. First, they expand topical coverage, particularly for underrepresented aspects such as service and ambiance, and lead to longer reviews while reducing overall review volume. Second, reviews become more textually complex and less readable, and they receive fewer helpfulness votes on average. Further analysis shows that the decline in perceived helpfulness is mitigated when review content remains concentrated on a dominant dimension, highlighting the importance of informational focus. Lastly, we find that the impacts are heterogeneous: less experienced users expand topical coverage and review length more significantly, whereas experienced users exhibit greater complexity and larger declines in perceived helpfulness. Our findings extend research on AI and UGC by highlighting a distinct mode of AI deployment, i.e., guiding human contributions rather than generating the content on their behalf, and by revealing its benefits and unintended consequences for platform design.

**Keywords:** Artificial intelligence (AI), Online reviews, User-generated content, AI-assisted writing, Topic nudging, Review helpfulness

# 1 INTRODUCTION

Online reviews are a central form of user-generated content (UGC) that significantly shape consumer decision-making, influence firms' reputation and sales, and constitute a major source of value for the platforms that host them (Chevalier & Mayzlin, 2006; Liu et al., 2006; Forman et al., 2008). For online review platforms, the key challenge has been not only to collect more user-generated content but also to encourage users to contribute reviews that are sufficiently informative, comprehensive, and useful to others. In other words, platforms face a dual managerial problem: how to stimulate contribution while improving the quality of what participants contribute. Prior research has shown that platforms can influence review generation through financial incentives, social norms, peer influence, editorial interventions, and platform recognitions, among others (e.g., Chen et al., 2010; Burtch et al., 2018; Khern-am-nuai et al., 2018; Ke et al., 2020; Deng et al., 2022; Jiang et al., 2023). However, most of these mechanisms operate either before users begin drafting or after content has been posted. As a result, platforms still have limited means to influence review quality at the moment of composition, when contributors are deciding what aspects of their experience to discuss and how much detail to provide.

Recent advances in artificial intelligence (AI), especially generative AI (GenAI) and related language technologies, have expanded the set of tools available for addressing the challenge. Most of the recent discussions have focused on AI as a substitute for human content production, such as systems that autonomously generate content (e.g., AI agents), or as a post-production aid, such as systems that summarize completed reviews for the audience (Berg et al., 2023; Su et al., 2024; Koo et al., 2025). Yet an equally important and increasingly relevant development is the use of AI to guide users during content generation. Rather than replacing contributors, AI can be embedded in the writing interface to provide prompts, structure, and feedback as users compose text (Dhillon et al., 2024).[1] This mode of deployment is important because it offers platforms an innovative way to improve UGC production while preserving human authorship. In settings such as online reviews, where authenticity and personal experience are central to the value of the content, platforms may prefer AI systems that guide users' writing rather than generate reviews on their behalf (Hadero, 2024).

[1] Examples include Google's writing assistants, Grammarly's assistant, and other tools with varying levels of AI scaffolding for co-writing. Typically, these tools either suggest text (next words, sentences, or even paragraphs) or check the grammar. Our focus, in contrast, is an AI tool that prompts topics to cover in the content.

Among these emerging interventions, AI-enabled topic nudging is especially consequential. A topic-nudging tool provides content contributors with real-time cues about dimensions they may want to address and can dynamically indicate whether those dimensions have already been covered. In this sense, it functions as AI-enabled scaffolding embedded within the composition interface. The importance of such a design extends beyond any single product design or platform.[2] Online review platforms, e-commerce marketplaces with user reviews, knowledge-sharing communities, and other digital platforms that rely on UGC increasingly face a similar design problem: how to encourage richer, more useful contributions without fully automating content production. AI-enabled topic nudging offers a scalable, low-friction solution to this problem because it can shape writing behavior in real time while leaving substantive control with the human contributor.

Despite the growing relevance of such AI-assisted writing tools, we know relatively little about how they affect the generation of online reviews in real-world environments. Existing research on AI and UGC has largely focused on settings in which AI substitutes for human contributions (Xue et al., 2023; Burtch et al., 2024; del Rio-Chanona et al., 2024; Li et al., 2024) or AI changes how users consume content after it has been created (Kim et al., 2024; Su et al., 2024; Cheng et al., 2025; Koo et al., 2025). A related stream examines chatbot-based review collection interfaces in controlled laboratory settings (Sachdeva et al., 2024). Much less is known, however, about AI that intervenes during the act of composition by guiding, rather than generating, content in a live platform setting. This is an important gap because AI embedded in the writing interface may reshape not only whether users contribute, but also what they write, how they allocate attention across topics, and how audiences perceive the resulting content.

[2] Similar design choices appear across many types of digital platforms. In this research, we examine Yelp that uses interactive review topics to help users write more informative and structured reviews in their own words. Other review platforms like Google Maps, Tripadvisor, and Airbnb use prompts (not all AI-powered) to elicit more informative first-hand reviews. Online communities adopt AI-powered tools, e.g., Reddit's Post Check, to help check users' post content. Knowledge-sharing platforms use structured forms and AI-assisted drafting with contributed content, e.g., Stack Overflow's Question Assistant and GitHub's Copilot-based issue creation tool. These examples suggest that platforms increasingly try to improve user contributions through prompts, structured fields, and light AI assistance rather than full automation.

The effects of such a tool are theoretically ambiguous. On the one hand, AI-enabled topic nudges may broaden topical coverage, reduce uncertainty about what to include, and encourage users to produce more informative and comprehensive reviews. On the other hand, the same feature may increase the effort required to complete each review, reallocate attention across multiple dimensions, and alter the narrative structure of the resulting content. A review that covers more aspects of an experience is not necessarily perceived as more helpful if the review becomes overly complex, less readable, or less focused on the most diagnostic dimension. In addition, the effects may differ systematically across users. Less experienced contributors may benefit more from such scaffolding because they lack established routines for writing reviews, whereas experienced contributors may find the same guidance redundant or even disruptive to their existing writing practices.

Motivated by the theoretical tensions, we ask the following questions. First, *how does AI-enabled topic nudging affect the generation of online reviews in both quantity and quality*? Second, *how do these effects vary across users with different levels of experience*?

We answer these questions by examining the introduction of a real-time, AI-enabled topic nudging tool into the review-composing interface on Yelp in 2023 as a unique empirical opportunity.[3] The feature provides AI-assisted writing support by prompting review contributors with key aspects of the dining experience, specifically "Food," "Service," and "Ambiance," during the review-writing process. In particular, the tool uses AI to dynamically signal whether each dimension has been covered as the user writes (i.e., highlighting the corresponding topic and marking it with a checkmark once it is covered). Figure 1 illustrates the format of Yelp's AI-enabled topic nudges. Since then, the design has been the default interface for writing restaurant reviews on Yelp.

We leverage Yelp's rollout of the feature as a natural experiment and employ a differences-in-differences approach (Han et al., 2022; Burtch et al., 2024; Eichenbaum et al., 2024; Goldberg et al.,

[3] We use the terms "AI-enabled topic nudges," "AI nudges," "topic nudges,", "the nudge" interchangeably in this research.

2024) that compares observations from calendar year 2023 (the inception of the tool) with the previous year across seven major cities in the United States.[4] Our dataset contains more than 1.5 million restaurant reviews across over 45,000 restaurants, which enables us to examine multiple dimensions of review generation at scale. In addition to review volume and length, we analyze topic distribution, textual complexity, readability, and helpfulness votes. We further explore heterogeneity by user experience and conduct multiple robustness checks, including alternative control groups and alternative specifications.

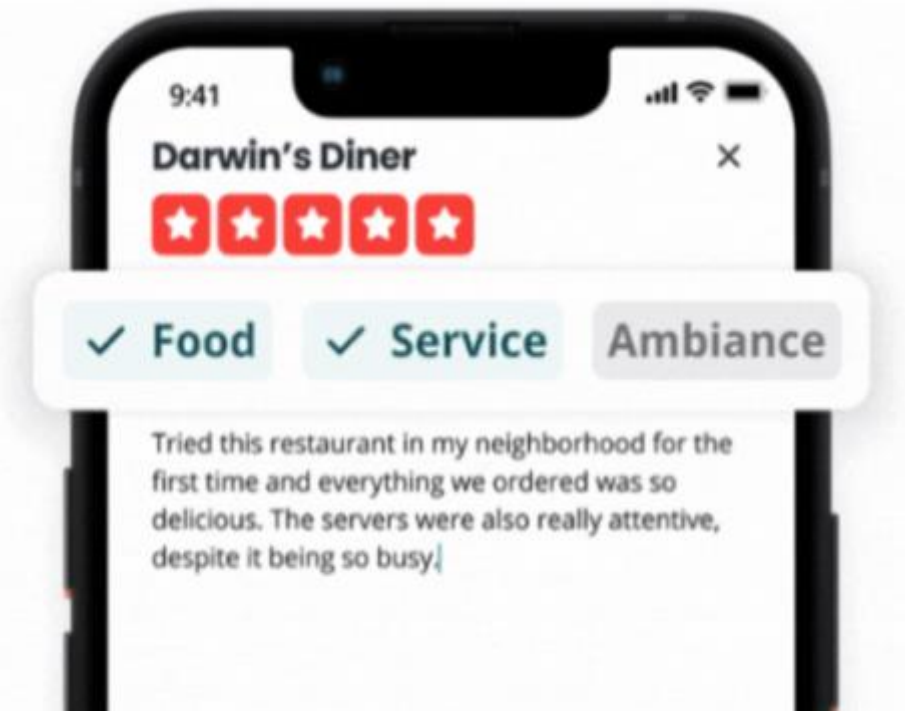

**Figure 1.** An Example of Yelp's Real-Time AI-Enabled Topic Nudges

Our findings reveal several important impacts of AI-enabled topic nudges on the generation of online reviews. First, the nudges reshape the topical composition of reviews by encouraging review writers to devote relatively greater attention to AI-suggested but traditionally underemphasized aspects of the dining experience, especially service and ambiance. This shift is accompanied by longer reviews, indicating that the tool successfully broadens the scope of discussion among users. At the same time, the number of reviews declines, suggesting a trade-off between the depth of individual contributions and the overall frequency of contributions. Second, although the AI-enabled nudges expand topical coverage, they also increase textual complexity and reduce readability. Consistent with these changes, reviews written after the introduction of the feature receive fewer helpfulness votes on average. Third, our mechanism analysis suggests that this decline in perceived helpfulness is attenuated when review content remains more concentrated on a dominant dimension, indicating that broader topic coverage may come at the cost of informational focus. Finally, the effects are heterogeneous across review writers: less experienced users respond more strongly in terms of review length and topic coverage, whereas experienced users experience a sharper decline in perceived helpfulness.

[4] Results are consistent when comparing Yelp restaurant reviews with (1) Google restaurant reviews or (2) Yelp reviews from another major category, both unaffected by AI-enabled topic nudging during our study period.

These findings contribute to at least three strands of the literature. First, we contribute to the literature on AI and UGC by identifying and analyzing a distinct mode of AI deployment: AI as a guide to human content production rather than a substitute for it. In this research, we move beyond the now-familiar contrast between human-generated and AI-generated content (Burtch et al., 2024; Li et al., 2024; Shankar et al., 2024; Cheng et al., 2025) and instead study a non-laboratory co-production setting in which AI shapes the writing process in real time (Sachdeva et al., 2024). Second, we contribute to the emerging literature on AI-assisted writing (Bhat et al., 2023; Dhillon et al., 2024) by showing that the effects of assistance depend not only on whether AI is present, but also on how that assistance is embedded in the interface and how users with different levels of experience respond to it. Third, we contribute to the literature on online review management by examining an intervention that simultaneously affects both review quantity and quality. Prior work has documented how monetary incentives, social influence, and editorial features shape review generation (Burtch et al., 2018; Khern-am-nuai et al., 2018; Wang et al., 2019a); our study shows that AI-enabled interface design is another important governance mechanism that introduces a nontrivial trade-off between broader content coverage and perceived helpfulness.

Our study also offers practical implications for platforms and content contributors. More broadly, our findings suggest that even light-touch AI scaffolding can meaningfully reshape human-generated content. For online review platforms, this means that AI-enabled topic nudging should not be treated merely as a convenience feature, but as a consequential governance mechanism that affects both content production and audience evaluation. Although such nudges can encourage richer and more balanced reviews, they may also increase writing costs and weaken the informational focus that readers value. Platforms therefore need to calibrate AI-assisted writing tools carefully so that they expand coverage without undermining clarity and usefulness. In addition, platforms should be attentive to heterogeneity across contributors. In our setting, less experienced users appear to benefit more from AI scaffolding, whereas experienced users experience sharper declines in perceived helpfulness. This suggests that a one-size-fits-all design is unlikely to be optimal; platforms may benefit from providing stronger guidance to novice users while making AI support lighter, optional, or less intrusive for experienced contributors.

## 2 LITERATURE AND CONTRIBUTIONS

This study builds on three strands of literature: the impact of AI on user-generated content, AI-assisted writing, and the management of online reviews.

### 2.1 Impacts of AI on User-Generated Content

Our study relates to a growing literature on how AI affects user-generated content. Recent work has focused primarily on how generative AI substitutes for user contributions in online knowledge-sharing communities. For example, the launch of large language models such as ChatGPT led to sharp declines in both questions and high-quality answers on platforms like Stack Overflow (Xue et al., 2023; Burtch et al., 2024; Borwankar et al., 2024; del Rio-Chanona et al., 2024; Li et al., 2024). Platform-level responses, such as banning AI-generated content, did not reverse the decline in question-asking (Wang & Zheng, 2024). In online reviews, prior work has primarily examined how AI-generated summaries affect subsequent user contributions. These summaries can increase contribution volume, attract more comments and unique viewers, make later reviews more textually similar to the summaries, and are also associated with higher ratings and lower rating dispersion (Su et al., 2024; Kim et al., 2024; Koo et al., 2025; Cheng et al., 2025). Other studies examine how AI affects review quality more directly. Knight et al. (2023) show that AI-detected reviews receive lower helpfulness votes on Yelp and Amazon. Sachdeva et al. (2024) compare chatbot-based review collection with traditional web forms through lab experiments and find that adding structured guidance, such as prompts about specific aspects of the experience, can improve review length and quality, but does not fully offset the negative effects of chatbots.[5]

We contribute to the literature in several ways. First, we extend the literature by examining a less-studied form of AI use: AI as a real-time nudge rather than a content substitute that generates new content or processes existing content. This highlights a distinct form of human–AI co-production in which AI shapes how users write rather than replacing them. Second, different from Sachdeva et al. (2024), the AI-enabled topic nudges we study provide dynamic writing support as users compose their reviews, rather than asking them to answer preset questions. This setting allows us to examine a wider range of outcomes, including topic coverage and helpfulness votes, test heterogeneity across users with different levels of prior platform experience, and explore the mechanism through which these nudges affect perceived review quality.

[5] In their study, review quality is measured using a composite index that combines participants' helpfulness ratings and a computed argument-quality score.

### 2.2 AI-Assisted Writing

The rapid development of LLMs has led to the widespread adoption of AI-assisted writing tools that support users throughout the writing process. Prior research shows that AI writing assistance can help with idea generation, narrative development, collaborative storytelling, and iterative refinement, thereby improving both writing productivity and the quality of written output (Kreminski et al., 2020; Lee et al., 2022; Singh et al., 2023).

Recent research further shows that AI-assisted writing can influence economic outcomes in contexts where written communication is central. For example, by helping users produce better-aligned text, AI tools can improve outcomes such as hiring and callback probabilities, while also lowering the effort required to participate in text-based activities in online labor markets (Wiles et al., 2025; Cui, Dias, and Ye, 2025; Galdin and Silbert, 2025). At the same time, AI-assisted writing may reduce the signaling value of written applications, making it harder for audiences to distinguish underlying quality (Galdin and Silbert, 2025; Chen and Chan, 2024). More importantly, the effects of AI-assisted writing could vary across users. Prior work largely confirms such observations: less experienced writers tend to benefit more from AI support, whereas expert writers may benefit less or even experience drawbacks when assistance is overly intrusive (Bhat et al., 2023; Petridis et al., 2023; Dhillon et al., 2024).

We contribute to this literature by showing that the effects of assistance depend not only on whether AI is present, but also on how that assistance is embedded in the interface and how users with different levels of experience respond to it. In our setting, AI is integrated into the review-writing interface through topic suggestions and real-time feedback, shaping the writing process as users compose their reviews. We show that this form of assistance affects not only how much users write, but also what they write about, and how those reviews are evaluated by audiences. We further document heterogeneous responses across users with different levels of prior experience. By linking interface design, writing behavior, and audience response, our study extends the AI-assisted writing literature to large-scale online review settings.

### 2.3 Management of Online Reviews

Research on managing the content on review platforms has developed along two primary sub-streams. The first focuses on motivating users to contribute more reviews, as greater review volume can improve consumer decision-making and sales (Chevalier & Mayzlin, 2006; Forman et al., 2008; Liu, 2006). Given that writing reviews resembles contributing to a public good (Bolton et al., 2004), review platforms have implemented various strategies to encourage participation. First, financial incentives have been shown to increase review volume, but sometimes at the cost of review quality, length, and even continued contributions (Burtch et al., 2018; Khern-am-nuai et al., 2018; Qiao et al., 2020; He et al., 2023). Second, social norms also play an important role: for example, highlighting users' median contribution levels (Chen et al., 2010) and displaying peers' contributions (Wang et al., 2019a; Ke et al., 2020) can encourage greater participation, and combining these cues with financial incentives can further increase both the number and length of reviews (Burtch et al., 2018).

The second sub-stream focuses on strategies to enhance the helpfulness and quality of reviews. More reviews do not necessarily mean more informative content (Mudambi & Schuff, 2010; Cao et al., 2011). In general, more detailed reviews are considered more helpful (Mudambi & Schuff, 2010), but overly complex language or excessive information can reduce usefulness and discourage engagement (Ghose et al., 2019; Yang et al., 2023). Emotional tone also matters: moderate emotionality tends to increase helpfulness (Yin et al., 2014), whereas strongly negative emotion can make reviews seem less helpful (Yin et al., 2021). Beyond review content itself, social and platform design features also shape review quality. For example, visible social cues can increase review length and novelty (Ke et al., 2020), recognition mechanisms such as awards can encourage more novel contributions (Burtch et al., 2022), editorial reviews can stimulate longer and more diverse user responses (Deng et al., 2022), and subscription features that signal popularity can indirectly improve readability (Goes et al., 2014)

We contribute to the literature by examining an intervention that simultaneously affects both review quantity and quality. Prior work has shown that monetary incentives, social influence, and editorial features shape review generation, but less is known about how AI embedded directly in the writing interface influences contribution behavior. We show that AI-enabled topic nudges affect how reviews are written and evaluated and introduce a meaningful trade-off between broader content coverage and perceived helpfulness. We further show that these effects vary across users with different experience levels, offering evidence on how platforms can use AI tools to govern content production.

## 3 HYPOTHESIS DEVELOPMENT

We conceptualize Yelp's AI-enabled topic nudges as a form of digital nudging, in which interface design elements guide user behavior without restricting choice (Weinmann et al., 2016; Mirsch et al., 2017; Schneider et al., 2018). Digital nudging has received growing attention in the IS literature as a way to influence user behavior through digital design.[6] In the Yelp review context, the nudge is embedded directly into the review-writing interface, providing real-time visual cues and completion signals that encourage users to address multiple dimensions of their experience, including food, service, and ambiance. Unlike static writing instructions or ex-ante guidelines, this intervention is embedded directly into the review-writing interface. As users compose a review, they continuously receive feedback about which dimensions have been addressed, creating a checklist-like environment that interacts with their ongoing writing process.

Prior research shows that digital nudges shape user behavior through interface-based cues during users' ongoing interactions with digital systems (Mirsch et al., 2017; Schneider et al., 2018; Dennis et al., 2020; Gupta et al., 2024). In our setting, the nudge can affect how they allocate attention and effort during content generation. It provides structural guidance that helps users organize their experiences, while simultaneously requiring them to monitor completion prompts during writing. As a result, the nudge can change both the effort required to produce reviews and the distribution of information across topics, thereby affecting both the quantity and perceived quality of user-generated reviews.

[6] See, for example, Hashim et al. (2017), Kretzer and Maedche (2018), Hashim et al. (2018), Wang et al. (2018), Dennis et al. (2020), Li et al. (2021), Xiao et al. (2022), Lee et al. (2023), Zeng et al. (2023), Gupta et al. (2024), Hwang and Lee (2025), Rivera et al. (2025), and Shukla et al. (2025), for recent research in information systems on digital nudging.

### 3.1 Production Trade-off: Cognitive Scaffolding and Effort Costs

By directing users' attention to predefined dimensions of their experience, Yelp's AI-enabled writing nudges provide a structural scaffold that helps them organize their reviews. Prior research on scaffolding suggests that external guidance can help individuals perform complex tasks by providing a framework that structures how the task should be completed (Wood et al., 1976). In the context of online reviews, the nudge helps users recall different aspects of their experience and organize them into a multi-dimensional narrative, which could lead to longer reviews and broader topical coverage.

At the same time, the interactive design of the nudges may increase the effort required to write a review. Because users receive real-time feedback about which dimensions have been addressed, they must split their attention between composing the review and monitoring the interface. Prior research in cognitive load theory shows that dividing attention across multiple information sources increases cognitive effort and creates a split-attention environment (Sweller, 1988; Chandler and Sweller, 1991). In this sense, although the nudges make review structure easier, the need to monitor the interface and reallocate attention could increase the effort required to produce each individual review. As a result, individuals may reduce the frequency of participation due to time and cognitive resource constraints (Zhang and Zhu, 2011; Burch et al., 2018). In other words, although AI nudges may broaden topical coverage, the added efforts may reduce the overall number of reviews contributed. Therefore, we hypothesize that:

***Hypothesis 1 (H1):*** *AI-enabled topic nudges will lead to longer reviews and broader topical coverage, but a lower overall volume of reviews.*

### 3.2 Quality Paradox: Topic Redistribution and Helpfulness

Yelp's AI-enabled topic nudges can also change how information is distributed within a review. By prompting users to address multiple aspects of their experience, the feature shifts attention away from the most important dimension and toward a broader set of topics. This broader coverage can increase the range of information in a review, but it may also reduce depth on the most informative aspect. Prior research shows that readers find reviews more helpful when they provide detailed, relevant information about product or service attributes (Jiang and Benbasat, 2007; Filieri et al., 2018; Mudambi and Schuff, 2010; Willemsen et al., 2011). When attention is spread across several dimensions, users may offer more generic comments on less important aspects, which lowers the review's diagnostic value.

In addition, the nudges may also disrupt the natural flow of the review. As users try to cover multiple topics in response to prompts, the review may become less coherent and more complex, making it more difficult for readers to extract useful information efficiently (Ghose and Ipeirotis, 2011; Goes et al., 2014). As a result, although the AI-enabled writing nudges expand topical coverage, the resulting reviews may be perceived as less helpful. Thus, we hypothesize that:

***Hypothesis 2 (H2):*** *AI-enabled topic nudges will lead to lower perceived helpfulness of reviews, increased textual complexity, and lower readability*.

### 3.3 Moderating Role of User Experience

The effects of AI-enabled writing nudges are also likely to depend on users' prior experience on the platform. Less experienced users often lack well-developed schemas and routines for writing reviews and therefore rely more on external guidance. Prior research on scaffolding suggests that structured guidance helps novices organize their actions and complete tasks more effectively (Wood et al., 1976). As a result, less experienced users may respond strongly to the nudges and exhibit greater changes in review length and topical coverage. Because they must both write the review and monitor the interface prompts, the added effort may further reduce their overall participation. In contrast, experienced users are more likely to have well-developed schemas and routines for describing their experiences (Chi et al., 1988), which makes them less dependent on external scaffolding when composing reviews. Research on the expertise reversal effect suggests that guidance that benefits novices may become redundant or even disruptive for more experienced individuals (Kalyuga, 2009). In this setting, the nudges may interfere with their established writing routines and introduce additional informational noise or textual complexity.

These differences may also affect how reviews are evaluated. Readers often hold higher expectations for reviews written by experienced users (Bhattacherjee, 2001), and review writers' characteristics can shape perceived helpfulness (Forman et al., 2008). As a result, increases in complexity or reductions in readability may be judged more critically for reviews written by experienced users, leading to a stronger decline in perceived helpfulness. Taken together, these mechanisms suggest that the AI-enabled topic nudge may have heterogeneous effects across users with different levels of experience. Thus, we hypothesize that:

***Hypothesis 3 (H3):*** *The effects of AI-enabled topic nudges on review length, topical coverage, and review volume will be stronger for less experienced users than for experienced users.*

***Hypothesis 4 (H4):*** *AI-enabled topic nudges will lead to a greater decline in perceived helpfulness for reviews written by experienced users than for those written by less experienced users.*

# 4 RESEARCH CONTEXT AND DATA

## 4.1 Yelp's AI-Enabled Topic Nudges

Our study examines Yelp, one of the largest user-generated review platforms in the United States. As of 2025, Yelp had hosted over 330 million reviews in the United States.[7] In April 2023, Yelp incorporated the so-called "real-time AI-enabled topic nudges" into its restaurant reviews, using AI models trained on its own data and third-party sources (Cheng, 2023). This feature was initially limited to restaurant reviews but was expanded to other categories in early 2024 (Saldanha, 2024).

Yelp's AI-enabled topic nudges are designed to assist review writers during the writing process by providing real-time feedback on the topical coverage of their reviews. The feature uses Yelp's in-house AI models to detect whether users have discussed key aspects of the dining experience—specifically "Food," "Service," and "Ambiance"—as they compose their text. When drafting a review, users can see these three topics displayed above the text box, which serve as prompts reminding them of aspects they may wish to address (Saldanha, 2023). As the user writes, AI analyzes the content and signals coverage of a topic by turning the corresponding label green and adding a checkmark in real time. In this way, the nudges not only prompt potential topics but also dynamically monitor the review text and provide immediate feedback on topic coverage. This design is now the default interface for Yelp restaurant reviews. The format of the real-time AI-enabled topic nudges is shown in Figure 1.

[7] See: https://www.yelp-press.com/company/fast-facts/default.aspx (retrieved March 26, 2026).

### 4.2 Data collection

We started collecting data in March 2024. Following Aneja et al. (2025), we cover reviews up to that date from seven major cities in the United States: Atlanta, Chicago, Houston, Los Angeles, Minneapolis, New York, and San Francisco. To ensure our analysis captures meaningful review activity, we focus on restaurants that maintain an active presence on the platform. Many restaurants receive few or no reviews over extended periods of time, making it difficult to assess trends in customer engagement and review dynamics. To mitigate this issue and ensure a consistent level of activity in our sample, we restrict our analysis to active restaurants with at least one review posted in 2022. Additionally, to account for holiday surges in review activity, we focus on reviews posted between the second week of January and the third week of November in both 2022 and 2023, excluding New Year's and Thanksgiving holidays. The final dataset includes 44,019 restaurants and 1,559,528 reviews. Table 1 provides the summary statistics of all variables used in our empirical analyses.

## 5 EMPIRICAL ANALYSIS AND FINDINGS

This section presents our empirical design and main findings. We first describe the empirical strategy in Section 5.1. We then examine how AI-enabled topic nudges affect topic distribution and review generation in Section 5.2 and perceived helpfulness in Section 5.3. Section 5.4 explores the mechanisms behind the changes in helpfulness, and Section 5.5 investigates heterogeneous effects by user experience. Finally, Section 5.6 reports a series of robustness checks.

### 5.1 Empirical Design

We follow the prior literature on online reviews and conduct our analyses at the business level (Deng et al., 2022; Jiang et al., 2023). Yelp's introduction of AI-enabled topic nudges serves as a cross-sectional shock affecting all restaurant reviews on the platform;[8] we therefore employ a panel difference estimator, as outlined by Goldberg et al. (2024), to compare review outcomes in 2023 to those in 2022. The panel-difference estimator is similar to a difference-in-differences approach and has been widely used in information systems studies (e.g., Han et al., 2022; Burtch et al., 2024; Eichenbaum et al., 2024). It relies on the assumption that outcome variables follow stable seasonal patterns from one year to the next. This assumption is plausible in our setting because Yelp review activity exhibits recurring patterns, such as weekday–weekend cycles and holiday fluctuations. We further examine this identifying assumption using a relative-time specification reported later in the paper.

**Table 1.** Summary Statistics

| Variables | Mean | Std. Dev. | Min | Max |
|---|---|---|---|---|
| **Aggregated Review Characteristics (Restaurant level)** | | | | |
| *Weekly Review Volume* | 1.663 | 1.520 | 1.000 | 122.000 |
| *Weekly Review Length* | 441.815 | 373.773 | 23.000 | 5,219.000 |
| *Weekly Review Helpfulness Votes* | 1.377 | 4.698 | 0.000 | 793.000 |
| **Textural Review Characteristics (Review level)** | | | | |
| *Length* | 450.044 | 431.638 | 23.000 | 5462.000 |
| $Food_{score}$ | 0.408 | 0.309 | 0.001 | 0.998 |
| $Service_{score}$ | 0.245 | 0.283 | 0.001 | 0.998 |
| $Ambiance_{score}$ | 0.347 | 0.307 | 0.001 | 0.990 |
| $Food_{length}$ | 195.451 | 263.362 | 2.578 | 5286.467 |
| $Service_{length}$ | 137.339 | 255.728 | 1.328 | 4880.271 |
| $Ambiance_{length}$ | 117.254 | 132.470 | 1.355 | 3131.243 |
| $TopicEntropy$ | 0.745 | 0.294 | 0.019 | 1.386 |
| *Gunning Fog Index* | 7.544 | 3.326 | 1.120 | 198.730 |
| *SMOG Index* | 6.776 | 3.405 | 0.000 | 24.500 |
| *Lexical Density* | 0.806 | 0.115 | 0.110 | 1.000 |

*Note: The average length is calculated as the average number of characters per review, based only on English-language reviews. Weekly review volume, length, and helpfulness votes are aggregated at the restaurant level. Topic scores and topic length (see Section 5.2) are derived using LDA (Blei et al., 2003), and topic entropy is used to measure textual complexity from LDA (see Equation 2 in Section 5.2). Readability is measured by the Gunning Fog Index, SMOG Index, and lexical density (see Section 5.4.1) at the review level.*

Specifically, we treat 2023 as the treatment year and 2022 as the control year and estimate the effects of AI topic nudges at the business-week level. To address potential differences in underlying time trends across the two years, we detrend the outcome variables using the two-step procedure in Goodman-Bacon

[8] Several studies have employed a Regression Discontinuity in Time (RDiT) design when the treatment affects all users, attributing any abrupt change in outcomes to the treatment (Kanuri et al. 2025). However, in restaurant reviews, seasonal spikes—such as increased activity in summer—may violate RDiT's key assumption that outcomes evolve smoothly over time in the absence of treatment.

(2021): we first estimate group-specific trends using only control units and treated units in the pre-treatment period and then subtract these estimated trends from the outcomes.

We also conduct several robustness checks to strengthen our identification. First, we show that the results remain consistent when using the original non-detrended outcomes. Second, to mitigate potential contamination from the launch of ChatGPT on November 30th, 2022, we restrict the sample to the period between the second week of January and the third week of November in both 2022 and 2023. Within this window, reviews in 2023 may be uniformly exposed to a broader AI-related shock, whereas reviews in 2022 are not. Because the nudges affect the 2023 sample throughout the study period rather than only after Yelp's intervention, the panel-difference design could difference out this common year-level shift. Third, we show that the findings are robust to alternative control groups, including Google restaurant reviews and Yelp reviews from another major category that was not exposed to the nudges during our study period. Fourth, to account for unobserved local dynamics, we also estimate specifications with city-specific time trends. Finally, although our main specification is at the business level, the results are qualitatively consistent when estimated at both the review level and the review-writer level; these additional analyses are reported in Section 5.6 and the Appendix.

To address potential skewness, we adopt log-linear models for several key outcome variables. To mitigate concerns about heteroscedasticity, we employ robust standard errors clustered at the restaurant level (Bertrand et al., 2004). Each observation corresponds to a restaurant, with time periods defined at the weekly level. Our main regression specification is:[9]

$$\log Y_{itw} = \beta_1 Treat_t * Post_w + \beta_2 Treat_t + \gamma' \mathbf{X}_{itw} + \delta_i + \lambda_w + \epsilon_{itw}, \quad (1)$$

where $Y_{itw}$ denotes the detrended aggregated review characteristics for restaurant *i* in calendar week *w* in year *t* (either 2023 or 2022). $Treat_t$ is a dummy indicating year 2023, and $Post_w$ is a dummy indicating weeks after April 25th (for both years). $\mathbf{X}_{itw}$ includes controls such as prior weekly cumulative review volume and length. $\delta_i$ and $\lambda_w$ are restaurant and calendar week fixed effects. The coefficient $\beta_1$ captures the impact of Yelp's AI-enabled topic nudges.

[9] As a specification test, we also assign the same restaurant across different years different identifiers and re-estimate the regressions. The results are consistent. We prefer our main specification because the same restaurant across different years remains the same economic unit, so using a common identifier preserves the within-restaurant comparison and allows restaurant fixed effects to absorb persistent heterogeneity.

## 5.2 Effects of AI-Enabled Topic Nudges on Topic Distribution and Review Generation

### 5.2.1 Effects of AI-Enabled Topic Nudges on Topic Distribution

The primary purpose of AI-enabled topic nudges is to help users generate more comprehensive, well-structured reviews by suggesting content dimensions to consider and providing real-time feedback (Saldanha, 2023). In line with this stated goal, our first step is to examine whether AI-enabled topic nudges are effective in shaping the topical composition of reviews. Specifically, we test whether the introduction of the nudge leads to broader coverage across different aspects of the dining experience, thereby expanding the distribution of topics addressed in Yelp reviews.

We employ Latent Dirichlet Allocation (LDA) (Blei et al., 2003) to analyze how topic coverage changes with the AI-enabled topic nudges. LDA is a prevalent method in the IS literature for analyzing online reviews (e.g., Khern-am-nuai et al., 2018; Deng et al., 2022). Following Blei et al. (2003), Mimno et al. (2011), and Zhao et al. (2015), we select four topics based on perplexity, rate of perplexity change (RPC), and coherence scores.[10] For each review, LDA assigns topic scores that represent the proportion of the review content attributed to each topic, with scores summing to 1, indicating a probability distribution.

Typically, topic labels are assigned manually (Mejia et al., 2021), which can introduce biases such as confirmation bias. Recent studies suggest that LLMs can effectively handle text annotation and labeling tasks (Gilardi et al., 2023). Thus, we selected the top 100 reviews per topic with scores exceeding 0.95 and used the GPT-4.0-turbo API to generate labels based on keywords and top documents.[11] Based on top words and reviews, GPT-generated labels are "Restaurant Service Complaints," "Restaurant Reviews and Dining Experience," "Café and Desserts," and "Restaurant Ambiance." These are aggregated into three main categories: *Service, Food (including Drink), and Ambiance.* However, the decrease in the proportion of topics does not necessarily indicate a decrease in the content length of such topics, since topic distributions in LDA are normalized. Therefore, following Deng et al. (2022), we multiply the review-level content length by the relative topic proportion (*Service Lengths, Food Lengths, Ambiance Lengths*) to further explore how topic distribution and information density change.

[10] The comprehensive evaluation of LDA is reported in Appendix A.1.
[11] Appendix A.2 presents the keywords used to interpret the semantic meaning of the extracted topics and reviews most heavily associated with four topics.

We first examine how AI-enabled topic nudges affect topic distribution by estimating Equation (1) using the three topic scores and topic-specific lengths. The results are presented in Table 2. Columns (1)-(3) show a clear reallocation of topical emphasis: after the introduction of the nudges, the service score increases significantly, while the food score decreases significantly. This pattern suggests that the nudges shift review content away from food—the traditionally dominant dimension—and toward service. At the same time, columns (4)-(6) show that topic-specific length increases for all three dimensions. Importantly, the increases are larger for service and ambiance than for food, indicating that the nudges do not simply crowd out food discussion, but instead expand discussion of the less dominant dimensions more strongly. Taken together, these results suggest that AI-enabled topic nudges broaden topical coverage and redistribute attention across dimensions by encouraging users to elaborate beyond the main aspect of the experience, consistent with Hypothesis H1.

**Table 2.** Effects of the AI-Enabled Topic Nudges on Topic Distribution

| Dep. Var.: | (1) log(*Service Score*) | (2) log(*Food Score*) | (3) log(*Ambiance Score*) | (4) log(*Service Lengths*) | (5) log(*Food Lengths*) | (6) log(*Ambiance Lengths*) |
|---|---|---|---|---|---|---|
| Treat*Post | 0.006*** | -0.005*** | -0.001 | 0.055*** | 0.020** | 0.033*** |
| | (0.001) | (0.001) | (0.001) | (0.008) | (0.006) | (0.006) |
| Treat | -0.000 | 0.0001 | -0.0002 | -0.010 | -0.007 | -0.005 |
| | (0.001) | (0.001) | (0.001) | (0.007) | (0.005) | (0.005) |
| Prior length | -0.000*** | -0.000*** | -0.000*** | -0.0004*** | -0.0002*** | -0.0001*** |
| | (0.000) | (0.000) | (0.000) | (0.000) | (0.000) | (0.000) |
| Prior volume | -0.003*** | -0.001*** | -0.005*** | -0.002*** | -0.001 | 0.022*** |
| | (0.0004) | (0.0004) | (0.0005) | (0.004) | (0.003) | (0.003) |
| Restaurant FE | Yes | Yes | Yes | Yes | Yes | Yes |
| Week FE | Yes | Yes | Yes | Yes | Yes | Yes |
| Num. Obs. | 847,902 | 847,902 | 847,902 | 847,902 | 847,902 | 847,902 |
| $R^2$ | 0.041 | 0.036 | 0.037 | 0.035 | 0.035 | 0.039 |

*Note: Standard errors in parentheses are robust and clustered at the restaurant level.* $*p < 0.1$, $**p < 0.05$, $***p < 0.01$.

### 5.2.2 Effects of AI-Enabled Topic Nudges on Review Generation

We further examine how the AI-enabled topic nudges shape review length and the number of reviews. Table 3 reports the regression results, which indicate that AI-enabled topic nudges significantly increased review length while decreasing review volume. Numerically, reviews written with the AI-enabled topic nudges were on average 4.8% longer,[12] while the number of reviews reduced by about 2.5%. Taken together, these findings support the idea that AI-enabled topic nudges encourage more elaboration within each review but may also increase the effort required to contribute, thereby lowering overall participation, consistent with hypothesis H2.

**Table 3.** Effects of the AI-enabled topic nudges on Review Generation

| | (1) | (2) |
|---|---|---|
| Dependent Var.: | log($Lengths$) | log($Review\ Volume$) |
| Treat*Post | 0.048*** | -0.025*** |
| | (0.004) | (0.002) |
| Treat | -0.006* | -0.007*** |
| | (0.003) | (0.001) |
| Prior length | -0.001*** | 0.000 |
| | (0.000) | (0.000) |
| Prior volume | -0.008** | 0.054*** |
| | (0.003) | (0.002) |
| Restaurant FE | Yes | Yes |
| Calendar Week FE | Yes | Yes |
| Num. Obs. | 847,902 | 847,902 |
| $R^2$ | 0.039 | 0.037 |

*Note: The average length includes only reviews written in the English language. Standard errors in parentheses are robust and clustered at the restaurant level.* $* p < 0.1$, $** p < 0.05$, $*** p < 0.01$.

The control variables also show intuitive patterns: prior cumulative review length is negatively associated with current review length, while prior cumulative review volume is negatively associated with review length but positively associated with current review volume. This suggests that businesses with more active review histories tend to receive more reviews, but those reviews are, on average, shorter.

### 5.3 Effects of AI-Enabled Topic Nudges on Perceived Helpfulness

The results in Section 5.2 show that AI-enabled topic nudges increase review length and broaden topic coverage. We next examine how this shift affects perceived helpfulness. As discussed in Section 3.2, greater elaboration does not necessarily make reviews more useful to readers. When attention is spread across multiple dimensions, reviews may become less focused and harder to process, reducing

[12] The results for review length remain consistent when measured by word count.

their perceived helpfulness.

To test this hypothesis, we estimate Equation (1) using helpfulness votes as the dependent variable. The results, reported in Table 4, show that reviews written with the AI-enabled topic nudges received about 3.7% fewer helpfulness votes on average, indicating that they were perceived as less helpful by readers. Consistent with our hypothesis H2, although the nudges broaden review content, they may also make reviews less helpful from the reader's perspective. We further explore the underlying mechanisms behind the decline of helpfulness in the next section.

**Table 4.** Effects of the AI-Enabled Topic Nudges on Helpfulness

| Dep. Var.: | log ($Helpfulness\ Votes$) |
|---|---|
| Treat*Post | -0.037*** |
| | (0.004) |
| Treat | -0.004 |
| | (0.003) |
| Prior length | -0.000*** |
| | (0.000) |
| Prior volume | -0.030*** |
| | (0.003) |
| Restaurant FE | Yes |
| Week FE | Yes |
| Num. Obs. | 847,902 |
| $R^2$ | 0.032 |

*Note. Standard errors in parentheses are robust and clustered at the restaurant level.* * $p < 0.1$, ** $p < 0.05$, *** $p < 0.01$.

### 5.4 Mechanism Behind the Decline in Helpfulness

This section explores the mechanisms underlying the decline in helpfulness votes. We examine two potential channels. First, we study whether the nudges increase textual complexity and reduce readability, making reviews harder for readers to process. Second, we examine whether the nudges change how information is distributed across topics by reducing concentration and weakening informational focus.

#### 5.4.1 The Role of Textual Complexity and Readability

We have shown that AI-enabled topic nudges lead users to write longer reviews that cover more aspects of their experiences. However, greater elaboration may also make reviews more complex and less coherent, making it harder for readers to extract useful information (Goes et al., 2014). As a result, perceived helpfulness may decline (Korfiatis et al., 2012; Ghose et al., 2019; Wang et al., 2019b; Yang et al., 2023). To examine this mechanism, we first analyze how AI-enabled topic nudges affect textual complexity and readability.

First, we use the entropy of the review's topic features derived from LDA to measure textual complexity, a widely used approach in prior IS research (Gong et al., 2018; Ghose et al., 2019; Yang et al., 2023). Specifically, topic entropy is defined as the Shannon index of the review's topic features:

$$TopicEntropy_j = -\sum_{k=1}^{K} p_{jk} \log(p_{jk}), \tag{2}$$

where $p_{jk}$ is the topic probability of review $j$ over topic $k$. Higher values of the topic entropy measure indicate greater content complexity.

Second, we examine readability, which reflects the cognitive effort required for users to understand a review. Following Goes et al. (2014), we adopt the Gunning Fog Index to measure readability. The Gunning Fog Index estimates the years of formal education required to understand a text, based on sentence length and the proportion of complex words (i.e., words with three or more syllables). Higher values indicate that a text requires more education to read fluently and therefore reflects lower readability.

**Table 5.** Effects of the AI-Enabled Topic Nudges on Entropy and Readability

| Dep. Var.: | (1)<br>log (*Entropy*) | (2)<br>log (*Gunning Fog Index*) |
|---|---|---|
| Treat*Post | 0.002** | 0.008*** |
| | (0.001) | (0.001) |
| Treat | 0.000 | -0.001 |
| | (0.000) | (0.001) |
| Prior length | -0.000*** | -0.000*** |
| | (0.000) | (0.000) |
| Prior volume | -0.0004*** | 0.000*** |
| | (0.0003) | (0.000) |
| Restaurant FE | Yes | Yes |
| Week FE | Yes | Yes |
| Num. Obs. | 847,902 | 847,902 |
| $R^2$ | 0.037 | 0.039 |

*Note. Standard errors in parentheses are robust and clustered at the restaurant level.* * $p < 0.1$, ** $p < 0.05$, *** $p < 0.01$.

The results are reported in Table 5. In column (1), the positive and significant coefficient on topic entropy suggests that AI-nudged reviews are more complex. Consistent with this pattern, in column (2), the readability results show that AI-nudged reviews have a higher Gunning Fog Index, indicating lower readability. Taken together, these findings suggest that the added complexity and lower readability introduced by AI-enabled topic nudges may create informational noise, making reviews harder to process and reducing their perceived helpfulness.

### 5.4.2 The Role of Topic Imbalance

Previous results show that AI-enabled nudges broaden topical coverage, yet reviews written with the feature receive fewer helpfulness votes on average. These results suggest that changes in the topical structure of reviews may help explain the decline in perceived helpfulness. In this section, we examine this mechanism by linking helpfulness to the distribution of topics within a review.

To capture how review content is distributed across different aspects of the dining experience, we construct a measure of topic imbalance at the review level. Using the topic scores for the three main dimensions (Food, Service, and Ambiance), we first compute the share of the dominant topic for each review $j$:

$$DominantShare_j = \max\left(Food_{Score_j}, Service_{Score_j}, Ambiance_{Score_j}\right). \quad (3)$$

We then identify the share of the second-largest topic for each review $j$, denoted as $SecondShare_j$. The topic imbalance measure is defined as:[13]

$$Imbalance_j = DominantShare_j - SecondShare_j \quad (4)$$

A higher value of $Imbalance_j$ indicates that review content is more concentrated on a single aspect of the experience, whereas a lower value indicates that discussion is more evenly distributed across multiple aspects. We then aggregate $Imbalance_j$ to the business-week level, $Imbalance_{iw}$, and estimate the following specification to examine whether the effect of AI-enabled topic nudges on helpfulness votes varies with the concentration of review topics:

$$Y_{iwt} = \beta_1 Treat_t + \beta_2 Imbalance_{iwt} + \beta_3 Treat_t * Post_w + \beta_4 Treat_t * Imbalance_{iwt} + \beta_5 Post_w * Imbalance_{iwt} + \beta_6 Treat_t * Post_w * Imbalance_{iwt} + \alpha_i + \delta_w + \varepsilon_{iwt} \quad (5)$$

where $Y_{iwt}$ denotes the log number of helpfulness votes received by reviews in restaurant *i* in calendar week $w$ in year $t$ (either 2023 or 2022). $Imbalance_{iwt}$ measured the concentration of review topics using Equation (4). The key coefficient is $\beta_6$, which captures how the impact of AI-enabled topic nudges varies with topic imbalance.

[13]The results remain consistent if we define $Imbalance$ using only the pre-treatment period.

The results are presented in Table 6. The triple interaction term is positive and statistically significant, implying that the decline in perceived helpfulness is attenuated when review content remains more concentrated on a dominant dimension. In other words, readers respond more favorably to reviews that focus more clearly on a single aspect of the experience, indicating that the broader topic coverage introduced by the nudges may come at the cost of informational focus.

**Table 6.** Topic Imbalance and the Decline in Helpfulness Votes

| Dep. Var.: | log ($Helpfulness\ Votes$) |
|---|---|
| Treat | -0.028*** |
| | (0.006) |
| Imbalance | -0.053*** |
| | (0.008) |
| Treat*Post | -0.069*** |
| | (0.007) |
| Treat*Imbalance | -0.060*** |
| | (0.010) |
| Post*Imbalance | -0.023 |
| | (0.008) |
| Treat*Post*Imbalance | 0.059*** |
| | (0.011) |
| Restaurant FE | Yes |
| Week FE | Yes |
| Num. Obs. | 847,902 |
| $R^2$ | 0.002 |

*Note. Standard errors in parentheses are robust and clustered at the restaurant level.* * $p < 0.1$, ** $p < 0.05$, *** $p < 0.01$

Overall, the mechanism evidence supports hypothesis H2 by suggesting that the broader coverage induced by AI-enabled topic nudges may come at a cost: reviews become harder to process and less informationally focused, which helps explain the decline in perceived helpfulness.

### 5.5 Heterogeneous Effects of User Experience

To assess the heterogeneous effects by user experience, we classify users as experienced or less experienced based on Yelp's "Elite User" badges (Janik, 2025). Because this badge is awarded annually to users who consistently contribute high-quality, high-volume reviews, it provides a useful proxy for prior platform experience. We define a user as experienced if they have earned at least one Elite badge, and as less experienced otherwise.[14] Based on this classification, we split the sample by user type, aggregate the data to the restaurant level, and estimate Equation (1) separately for each group. Our results are robust to alternative definitions of experienced users, such as classifying users with at least one Elite badge since 2022 or since 2015 as experienced (see Appendix B).

The results are reported in Tables 7 and 8. Table 7 shows that less experienced users (Panel A) respond more strongly to the nudges in terms of topic allocation. Their reviews shift significantly toward service and away from food, and topic-specific length increases mainly for service and ambiance. In contrast, experienced users (Panel B) exhibit smaller changes in topic scores, although they still expand discussion length for some dimensions, especially ambiance. Columns (1) and (2) in Table 8 show a similar pattern for review generation. Less experienced users (Panel A) increase review length more and reduce review volume more than experienced users (Panel B). Taken together, these findings provide strong support for hypothesis H3, showing that prior platform experience shapes how users respond to AI-enabled topic nudges and that less experienced users respond more strongly in terms of review length and topic coverage.

**Table 7.** Heterogeneous Effects by User Experience on Topic Distribution

| Dep. Var.: | (1) log(*Service Score*) | (2) log(*Food Score*) | (3) log(*Ambiance Score*) | (4) log(*Service Lengths*) | (5) log(*Food Lengths*) | (6) log(*Ambiance Lengths*) |
|---|---|---|---|---|---|---|
| Panel A: Less Experienced Users | | | | | | |
| Treat*Post | 0.009*** | -0.008*** | -0.001 | 0.074*** | 0.004 | 0.028*** |
| | (0.001) | (0.001) | (0.001) | (0.009) | (0.007) | (0.007) |
| Treat | -0.000 | -0.000 | -0.000 | -0.006 | -0.004 | -0.005 |
| | (0.000) | (0.000) | (0.000) | (0.008) | (0.006) | (0.006) |
| Num. Obs. | 683,776 | 683,776 | 683,776 | 683,776 | 683,776 | 683,776 |
| Panel B: Experienced Users | | | | | | |
| Treat*Post | -0.001 | -0.003** | 0.004*** | 0.018 | 0.024** | 0.062*** |
| | (0.001) | (0.001) | (0.001) | (0.014) | (0.009) | (0.011) |
| Treat | -0.000 | 0.0003 | 0.000 | -0.011 | -0.004 | -0.003 |
| | (0.001) | (0.001) | (0.001) | (0.012) | (0.008) | (0.009) |
| Num. Obs. | 293,448 | 293,448 | 293,448 | 293,448 | 293,448 | 293,448 |
| Control Variables, restaurant FE, and calendar week FE included | | | | | | |

*Note: Standard errors in parentheses are robust and clustered at the restaurant level.* * $p < 0.1$, ** $p < 0.05$, *** $p < 0.01$.

[14] Previous research has shown that even when a review writer has been demoted from Elite status, consumers can still easily identify that they previously held the Elite status. Reviews written by demoted review writers are perceived by platform users to be of similar quality to those written by current Elite users (Pamuru et al. 2023).

**Table 8.** Heterogeneous Effects by User Experience on Review Generation

| Dependent Var.: | (1) log( $Lengths$) | (2) log( $Review\ Volume$) | (3) log( $Helpfulness\ Votes$) | (4) log($Entropy$) | (5) log ($Gunning\ Fog\ Index$) |
|---|---|---|---|---|---|
| Panel A: Less Experienced Users | | | | | |
| Treat*Post | 0.046*** | -0.022*** | -0.023*** | 0.001 | 0.008*** |
| | (0.004) | (0.002) | (0.003) | (0.001) | (0.002) |
| Treat | -0.004 | -0.005*** | -0.002 | -0.000 | -0.0005 |
| | (0.003) | (0.001) | (0.003) | (0.000) | (0.001) |
| Num. Obs. | 683,776 | 683,776 | 683,776 | 683,776 | 683,776 |
| Panel B: Experienced Users | | | | | |
| Treat*Post | 0.028*** | -0.008*** | -0.072*** | 0.003*** | 0.007*** |
| | (0.005) | (0.002) | (0.007) | (0.001) | (0.002) |
| Treat | -0.005 | -0.002 | -0.001 | -0.000 | -0.0007 |
| | (0.005) | (0.002) | (0.006) | (0.001) | (0.002) |
| Num. Obs. | 293,448 | 293,448 | 293,448 | 293,448 | 293,448 |
| Control Variables, restaurant FE, and calendar week FE included | | | | | |

*Note: The average length includes only reviews in the English language. Standard errors in parentheses are robust and clustered at the restaurant level.* $*p < 0.1$, $**p < 0.05$, $***p < 0.01$.

The two groups also differ in how their reviews are evaluated. Columns (3)-(5) of Table 8 show that, for less experienced users (Panel A), the nudges lead to a modest decline in helpfulness, with little change in textual complexity and a slight reduction in readability. In contrast, for experienced users (Panel B), the decline in perceived helpfulness is much larger—more than three times that of less experienced users—and is accompanied by significant increases in entropy and the Gunning Fog Index, indicating greater complexity and lower readability. Since readers hold higher baseline expectations for reviews from experienced contributors (Bhattacherjee, 2001), these changes may be judged more harshly. Taken together, these results are consistent with hypothesis H4, which posits that AI-enabled topic nudges reduced perceived helpfulness more for experienced users than for less experienced users.

## 5.6 Robustness Checks

### 5.6.1 A Relative-Time Model

To verify the parallel trends assumption for usual DID methods, we estimate a relative time model using leads and lags centered on the treatment week, following Angrist and Pischke (2009). The specification of our relative time model for the aggregated review characteristics analysis is:

$$\log Y_{itw} = \sum_j l_w Pre_{iw}(j) + \sum_l \tau_w Post_{iw}(l) + + \beta_2 Treat_t + g'X_{itw} + \alpha_i + \delta_w + \varepsilon_{itw}, \quad (6)$$

where $Y_{itw}$ represents one of the review characteristics (review volume, length, helpfulness votes, etc.) for restaurant $i$ in calendar week $w$ and year $t$. $\alpha_i$ and $\delta_w$ denote restaurant fixed effects and calendar week fixed effects, respectively. The newly added term $Pre_{iw}(j)$ is an indicator function that equals one if calendar week $w$ is $j$ week(s) prior to the treatment. Similarly, the term $Post_{iw}(l)$ is an indicator function that equals one if calendar week $w$ is $l$ week(s) after the introduction of AI-enabled topic nudges. To estimate the dynamic effects, we aggregate all pretreatment periods occurring at least eight weeks before treatment into a single indicator variable. Similarly, all posttreatment periods beginning eight weeks or more after treatment are combined into another indicator. We normalize the coefficient for the period immediately preceding treatment, $Pre_{iw}(-1)$, to zero, which serves as the baseline for comparison.

As shown in Appendix C, most coefficients are not statistically significant in pre-treatment periods. In particular, only six out of 70 coefficients (for the pretreatment dummies) are statistically sound at a significance level of 95% or higher. Such evidence is consistent with the assumption of parallel trends, lending further support to a causal interpretation of our estimates.

### 5.6.2 Alternative "Control" Groups

A potential concern in our main analysis is that the observed changes in Yelp restaurant reviews may reflect broader time-varying shocks in online review activity, rather than the introduction of AI-enabled topic nudges. To further address this concern, we conduct two additional robustness checks using alternative control groups.

First, we use the Beauty category on Yelp as an alternative within-platform benchmark. Beauty is the largest Yelp category after restaurants and similarly involves customer-facing services, but AI-enabled topic nudges were not introduced in its review interface during our study period. This makes Beauty reviews a useful comparison group for isolating the effect of the nudges from broader platform-level changes in Yelp review activity between 2022 and 2023. Because reviews in different service categories often emphasize different consumption dimensions, cross-category comparisons of topic distributions or other textual features may be difficult to interpret. We therefore focus only on aggregate review outcomes, including review volume, helpfulness votes, and mean review length.

Following our main identification strategy, we estimate a difference-in-difference-in-differences (DDD) specification that compares the change in restaurant reviews before and after the introduction of AI-enabled topic nudges with the corresponding change in Beauty reviews over the same period, while also differencing out year-level variation between 2022 and 2023. We estimate:

$$\log Y_{iwy} = \beta_1 Year2023_y + \beta_2 Treat_i * Post_w + \beta_3 Treat_i * Year2023_y + \beta_4 Post_w * Year2023_y + \beta_5 Treat_i * Post_w * Year2023_y + \alpha_i + \delta_w + \varepsilon_{iwy}, \quad (9)$$

where $Y_{iwy}$ denotes the review outcome for business $i$ in week $w$and year $y$. $Treat_i$ is an indicator equal to one if restaurant $i$ belongs to the restaurant category and zero if it belongs to the Beauty category. $Post_w$ is an indicator equal to one for weeks after the introduction of Yelp's AI-enabled topic nudges, and $Year2023_y$ is an indicator for observations in 2023. The coefficient of interest is $\beta_5$, which captures the policy effect in the DDD specification. This approach helps account for time-varying shocks affecting review activity across Yelp categories but unrelated to the nudges. The estimation results, reported in Appendix D, are consistent with our main findings: AI-enabled topic nudges significantly reduce review volume and helpfulness votes, while the effect on mean review length is statistically insignificant but directionally similar to the main estimates.

Second, we use Google reviews for the same set of restaurants as an alternative cross-platform control group. The intuition is that Yelp's AI-enabled topic nudges should affect review activity on Yelp but have little direct effect on reviews of the same restaurants on Google Maps. To ensure consistency, we restrict this analysis to the period after January 31, 2023, when Google introduced new review fields, and therefore focus on the first week of February through the third week of November 2023. Because Google allows users to leave ratings without accompanying text, potentially creating systematic differences in textual measures across platforms, we focus only on review volume and helpfulness votes in this analysis.

Following Jiang et al. (2023), we use a cross-platform DID identification strategy that compares changes in a restaurant's Yelp reviews following the launch of AI-enabled topic nudges with changes in the same restaurant's Google reviews over the same period. We estimate:

$$\log Y_{ipw} = \beta_1 Treat_p * Post_w + \alpha_{ip} + \delta_w + \varepsilon_{ipw}, \quad (7)$$

where $Y_{itw}$ denotes the review outcome for restaurant $i$ in week $w$ on platform $t$. $Treat_t$ is an indicator for reviews on Yelp and $Post_w$ is an indicator variable for weeks after Yelp's AI-enabled topic nudges. This strategy allows us to control for the unobserved shocks to restaurant reviews (e.g., improvements in service quality) that are common across both platforms. The results, reported in Appendix D, are again consistent with our baseline findings.

Taken together, these two robustness checks provide additional support for our main analysis and suggest that the estimated effects are unlikely to be driven by broader temporal shocks in review activity unrelated to Yelp's AI-enabled topic nudges.

#### 5.6.3 Additional Robustness Checks

We conduct several additional robustness checks to assess the robustness of our findings. First, we revisit the helpfulness-vote analysis using updated vote counts collected in January 2025. Because helpfulness votes are cumulative and non-decreasing over time, older reviews may mechanically receive more votes simply because they have been visible for longer. Our original dataset does not contain historical vote trajectories, which limits our ability to fully account for this issue. To address this concern, we re-collected helpfulness vote data in January 2025. Reviews from restaurants that were no longer available on Yelp at that time were excluded. Given the stability of helpfulness votes by 2025, we estimate Equation (1) to explore the impact of AI nudges on the updated vote counts. The results in Appendix E remain consistent with our main analysis.

Second, although our main analysis is conducted at the restaurant level, we also examine the effects of the AI-enabled topic nudges at both the review and review-writer levels. The results are reported in Appendix F and are qualitatively consistent with the main findings.

Third, our baseline analysis follows a two-step procedure, as in Goodman-Bacon (2021), and uses detrended outcome variables. As a robustness check, we re-estimate Equation (1) using the original, non-detrended outcomes. In addition, to better account for unobserved, city-level dynamics—particularly those stemming from differential recovery trajectories following the COVID-19 pandemic—we extend our DID analysis by incorporating city-specific time trends to Equation (1). The estimation results, reported in Appendix G, align with our main analysis and support the robustness of our findings.

# 6 CONCLUDING REMARKS

## 6.1 Summary

In this study, we examine how AI-enabled topic nudges embedded in a live writing interface shape the generation of online reviews. Rather than replacing human authors, these nudges guide users during composition by prompting them to address key aspects of the restaurant experience. Using Yelp's 2023 rollout as a natural experiment, we compare restaurant reviews from 2023 with those from 2022 across seven major U.S. cities.

Our findings yield three main takeaways. First, AI assistance can meaningfully reshape human-generated content even without generating the content itself. In our setting, this creates a production trade-off: reviews become longer and more broadly in topic coverage, but users contribute less frequently. Second, it creates an evaluation trade-off: although the nudges encourage more comprehensive reviews, the resulting content is perceived as less helpful by readers because broader coverage comes with greater complexity, lower readability, and a weaker informational focus. Third, the effects of AI assistance are heterogeneous across users: less experienced users appear to benefit more from its scaffolding role, whereas more experienced users appear more vulnerable to its disruptive effects.

Overall, our results show that AI-enabled topic nudges are a double-edged governance tool. They can encourage richer, more comprehensive reviews without replacing human authorship, but they can also reduce the frequency of contributions and lower perceived usefulness. These findings provide new insights into how AI-enabled interface features reshape user-generated content and highlight the importance of carefully designing AI-assisted writing tools on digital platforms.

## 6.2 Theoretical and Practical Implications

This study offers several theoretical contributions. First, it highlights a distinct role of AI in user-generated content: not as a substitute for human production, but as a writing guide embedded in the composition process. Second, it advances research on AI-assisted writing by showing that writing assistance operates through attention allocation during composition. Rather than simply helping users produce more content, AI assistance can redirect attention across multiple dimensions of a task, creating a tension between scaffolding and focus. Third, it identifies AI-enabled interface design as a platform governance mechanism that shapes not only how much users contribute but also what they write about and how their contributions are evaluated by readers.

From a practical perspective, our findings suggest that platforms should think about AI integration in content generation as a spectrum rather than a binary choice. At one end, AI can largely automate content production; in the middle, humans and AI co-produce content; and near the other end, humans remain the primary authors while AI plays only a light scaffolding role. Our setting lies in this last category. A key implication is that even minimal AI involvement can meaningfully reshape human-generated content. Platforms therefore should not assume that light-touch AI features are innocuous or uniformly beneficial simply because they do not generate content directly.

For online review platforms, AI-enabled topic nudging should be treated as a consequential governance mechanism rather than a mere convenience feature. Although the tool broadens topical coverage and lengthens reviews, it also reduces review volume and lowers perceived helpfulness by increasing complexity and weakening informational focus. The managerial challenge, therefore, is to calibrate AI assistance so that it expands coverage without undermining clarity and focus. One possible approach is to provide delayed or adaptive feedback. Rather than continuously displaying multiple topic prompts during writing, a platform could allow users to draft freely and then provide selective reminders afterward, such as suggesting one or two overlooked dimensions based on the content already written. This may preserve the user's natural flow while still encouraging broader coverage.

Finally, platforms should be attentive to heterogeneity. In our setting, less experienced users appear to benefit more from AI scaffolding, whereas experienced users experience a sharper decline in perceived helpfulness. This suggests that a one-size-fits-all AI design is unlikely to be optimal. Platforms may benefit from offering stronger guidance to novice users while making AI support lighter, optional, or less intrusive for experienced contributors. Such heterogeneity also matters strategically: if recognized contributors such as Yelp Elite users systematically suffer declines in helpfulness, the platform may discourage valuable contributors and weaken the credibility of its recognition system.

### 6.3 Limitations and Future Extensions

Our study can be extended in several ways for future research. First, our analysis focuses on the immediate, post-adoption phase, capturing user behaviors and textual outcomes soon after the AI-enabled topic nudges were introduced on Yelp. A longitudinal approach over an extended period could clarify whether the observed effect persists, diminishes, or intensifies with continued use of AI tools. Second, while we study online restaurant reviews, future research could explore other types of online content—for example, service or product reviews, online forum discussions, or social media posts—to examine whether AI-enabled topic nudges influence content production differently across contexts. Third, future work could also investigate other AI-powered technologies beyond AI-enabled topic nudges, such as AI-driven editing assistants, summarization tools, or image–text generation features, to better understand how these technologies shape user contributions.

# REFERENCES

Angrist, J. D., & Pischke, J. S. (2009). *Mostly harmless econometrics: An empiricist's companion*. Princeton University Press.

Aneja, A., Luca, M., & Reshef, O. (2025). The benefits of revealing race: Evidence from minority-owned local businesses. *American Economic Review, 115*(2), 660–689.

Bertrand, M., Duflo, E., & Mullainathan, S. (2004). How much should we trust differences-in-differences estimates? *Quarterly Journal of Economics, 119*(1), 249–275.

Berg, J. M., Raj, M., & Seamans, R. (2023). Capturing value from artificial intelligence. *Academy of Management Discoveries, 9*(4), 424–428.

Bhat, A., Agashe, S., Oberoi, P., Mohile, N., Jangir, R., & Joshi, A. (2023, March). Interacting with next-phrase suggestions: How suggestion systems aid and influence the cognitive processes of writing. In *Proceedings of the 28th International Conference on Intelligent User Interfaces* (pp. 436–452). Association for Computing Machinery.

Bhattacherjee, A. (2001). Understanding information systems continuance: An expectation-confirmation model. *MIS Quarterly*, 25(3), 351–370.

Blei, D. M., Ng, A. Y., & Jordan, M. I. (2003). Latent Dirichlet allocation. *Journal of Machine Learning Research, 3*(1), 993–1022.

Burtch, G., Hong, Y., Bapna, R., & Griskevicius, V. (2018). Stimulating online reviews by combining financial incentives and social norms. *Management Science, 64*(5), 2065–2082.

Burtch, G., He, Q., Hong, Y., & Lee, D. (2022). How do peer awards motivate creative content? Experimental evidence from Reddit. *Management Science, 68*(5), 3488–3506.

Burtch, G., Lee, D., & Chen, Z. (2024). The consequences of generative AI for online knowledge communities. *Scientific Reports, 14*(1), 10413.

Bolton, G. E., Katok, E., & Ockenfels, A. (2004). How effective are electronic reputation mechanisms? An experimental investigation. *Management Science, 50*(11), 1587–1602.

Borwankar, S., Khern-am-nuai, W., & Kannan, K. N. (2024). Unraveling the impact: An empirical investigation of ChatGPT's exclusion from Stack Overflow. *Available at SSRN.*

Cao, Q., Duan, W., & Gan, Q. (2011). Exploring determinants of voting for the "helpfulness" of online user reviews: A text mining approach. *Decision Support Systems, 50*(2), 511–521.

Chandler, P., & Sweller, J. (1991). Cognitive load theory and the format of instruction. *Cognition and Instruction*, 8(4), 293-332.

Chevalier, J. A., & Mayzlin, D. (2006). The effect of word of mouth on sales: Online book reviews. *Journal of Marketing Research, 43*(3), 345–354.

Chen, Y., Harper, M., Konstan, J., & Li, S. X. (2010). Social comparisons and contributions to online communities: A field experiment on Movie Lens. *American Economic Review, 100*(4), 1358–1398.

Chen, Z., & Chan, J. (2024). Large language model in creative work: The role of collaboration modality and user expertise. *Management Science, 70*(12), 9101–9117.

Cheng, M. (2023). "Yelp is using AI to help users write reviews." *Quartz*, https://qz.com/yelp-is-using-ai-to-help-users-write-reviews-1850370277. Accessed March 26, 2026.

Cheng, Y., Gao, B., Zhang, R. A., & Li, X. (2025). "Word-of-AI" and matching quality: Evidence from a natural experiment on online review platforms. *HEC Paris Research Paper No. MOSI-2025-1545, Available at SSRN.*

Chi, M. T., Glaser, R., & Farr, M. J. (2014). The nature of expertise. Psychology Press.

Cui, J., Dias, G., & Ye, J. (2025). Signaling in the Age of AI: Evidence from Cover Letters. arXiv preprint arXiv:2509.25054.

del Rio-Chanona, R. M., Laurentsyeva, N., & Wachs, J. (2024). Large language models reduce public knowledge sharing on online Q&A platforms. *PNAS Nexus, 3*(9), pgae400.

Deng, Y., Zheng, J., Khern-am-nuai, W., & Kannan, K. (2022). More than the quantity: The value of editorial reviews for a user-generated content platform. *Management Science, 68*(9), 6865–6888.

Dennis, A. R., Yuan, L., Feng, X., Webb, E., & Hsieh, C. J. (2020). Digital nudging: Numeric and semantic priming in e-commerce. *Journal of Management Information Systems*, 37(1), 39-65.

Dhillon, P. S., Molaei, S., Li, J., Golub, M., Zheng, S., & Robert, L. P. (2024, May). Shaping human-AI collaboration: Varied scaffolding levels in co-writing with language models. In *Proceedings of the CHI Conference on Human Factors in Computing Systems* (pp. 1–18). Association for Computing Machinery.

Eichenbaum, M., Godinho de Matos, M., Lima, F., Rebelo, S., & Trabandt, M. (2024). Expectations, infections, and economic activity. *Journal of Political Economy, 132*(8), 2571–2611.

Filieri, R., Hofacker, C. F., & Alguezaui, S. (2018). What makes information in online consumer reviews diagnostic over time? The role of review relevancy, factuality, currency, source credibility and ranking score. *Computers in Human Behavior*, 80, 122-131.

Forman, C., Ghose, A., & Wiesenfeld, B. (2008). Examining the relationship between reviews and sales: The role of reviewer identity disclosure in electronic markets. *Information Systems Research, 19*(3), 291–313.

Galdin, A., & Silbert, J. (2025). Making Talk Cheap: Generative AI and Labor Market Signaling. arXiv preprint arXiv:2511.08785.

Ghose, A., & Ipeirotis, P. G. (2010). Estimating the helpfulness and economic impact of product reviews: Mining text and reviewer characteristics. *IEEE transactions on knowledge and data engineering*, 23(10), 1498-1512.

Ghose, A., Ipeirotis, P. G., & Li, B. (2019). Modeling consumer footprints on search engines: An interplay with social media. *Management Science, 65*(3), 1363–1385.

Gilardi, F., Alizadeh, M., & Kubli, M. (2023). ChatGPT outperforms crowd workers for text-annotation tasks. *Proceedings of the National Academy of Sciences, 120*(30), e2305016120.

Goes, P. B., Lin, M., & Au Yeung, C. M. (2014). "Popularity effect" in user-generated content: Evidence from online product reviews. *Information Systems Research, 25*(2), 222–238.

Goodman-Bacon, A. (2021). Difference-in-differences with variation in treatment timing. *Journal of Econometrics, 225*(2), 254–277.

Goldberg, S. G., Johnson, G. A., & Shriver, S. K. (2024). Regulating privacy online: An economic evaluation of the GDPR. *American Economic Journal: Economic Policy, 16*(1), 325–358.

Gong, J., Abhishek, V., & Li, B. (2018). Examining the impact of keyword ambiguity on search advertising performance. *MIS Quarterly, 42*(3), 805–A14.

Gupta, P., Kim, Y. J., Glikson, E., & Woolley, A. W. (2024). Using digital nudges to enhance collective intelligence in online collaboration: Insights from unexpected outcomes. *MIS Quarterly*, 48(1), 393-408.

Hadero H (2024) The internet is filled with fake reviews. Here are some ways to spot them. AP News (December 23), https://apnews.com/article/fake-online-reviews-generative-ai-40f5000346b1894a778434ba295a0496.

Han, B. R., Sun, T., Chu, L. Y., & Wu, L. (2022). COVID-19 and E-commerce Operations: Evidence from Alibaba. *Manufacturing & Service Operations Management*, 24(3), 1388-1405.

Hashim, M. J., Kannan, K. N., & Maximiano, S. (2017). Information feedback, targeting, and coordination: An experimental study. *Information Systems Research*, 28(2), 289-308.

Hashim, M. J., Kannan, K. N., & Wegener, D. T. (2018). Central role of moral obligations in determining intentions to engage in digital piracy. *Journal of Management Information Systems*, 35(3), 934-963.

He, L., Luo, J., Tang, Y., Wu, Z., & Zhang, H. (2023). Motivating user-generated content: The unintended consequences of incentive thresholds. *MIS Quarterly, 47*(3), 1015–1044.

Hwang, E. H., & Lee, S. (2025). A nudge to credible information as a countermeasure to misinformation: Evidence from twitter. *Information Systems Research*, 36(1), 621-636.

Janik, A. (2025). "Welcome to the Elite Squad: FAQs." *Yelp*, https://blog.yelp.com/community/welcome-to-the-elite-squad-faqs/. Accessed March 26, 2026.

Jiang, Z., & Benbasat, I. (2007). The Effects of Presentation Formats and Task Complexity on Online Consumers' Product Understanding1. *MIS quarterly*, 31(3), 475-500.

Jiang, L. D., Ravichandran, T., & Kuruzovich, J. (2023). Review moderation transparency and online reviews: Evidence from a natural experiment. *MIS Quarterly*, 49(1), 275-304.

Kalyuga, S. (2009). The expertise reversal effect. In Managing cognitive load in adaptive multimedia learning (pp. 58-80). IGI Global Scientific Publishing.

Kanuri, V. K., Crecelius, A. T., & Kumar, S. (2025). Disentangling the Customer-Level, Cross-Channel Effects of Large-Order-Advantaged Online Shipping Policies. *MIS Quarterly*, 49(1), 275-304.

Ke, Z., Liu, D., & Brass, D. J. (2020). Do online friends bring out the best in us? The effect of friend contributions on online review provision. *Information Systems Research*, 31(4), 1322-1336.

Khern-am-nuai, W., Kannan, K., & Ghasemkhani, H. (2018). Extrinsic versus intrinsic rewards for contributing reviews in an online platform. *Information Systems Research*, 29(4), 871-892.

Kim, A., Lu, Y., Ma, T., & Tan, Y. (2024). Less is more? Impact of AI-generated summaries on user engagement of video-sharing platforms. *Available at SSRN.*

Knight, S., Bart, Y., & Yang, M. (2023). Generative AI and user-generated content: Evidence from online reviews. *Northeastern U. D'Amore-McKim School of Business Research Paper No. 4621982,*

Korfiatis, N., García-Bariocanal, E., & Sánchez-Alonso, S. (2012). Evaluating content quality and helpfulness of online product reviews: The interplay of review helpfulness vs. review content. *Electronic Commerce Research and Applications*, 11(3), 205-217.

Koo, W. W., Pu, J., & Gao, N. (2025). AI Review Summary, Customer Ratings, and Performance Entrenchment.

Kreminski, M., Dickinson, M., Mateas, M., & Wardrip-Fruin, N. (2020, September). Why Are We Like This? The AI architecture of a co-creative storytelling game. *In Proceedings of the 15th International Conference on the Foundations of Digital Games* (pp. 1-4).

Kretzer, M., & Maedche, A. (2018). Designing social nudges for enterprise recommendation agents: An investigation in the business intelligence systems context. *Journal of the Association for Information Systems*, 19(12), 4.

Lee, M., Liang, P., & Yang, Q. (2022, April). Coauthor: Designing a human-ai collaborative writing dataset for exploring language model capabilities. *In Proceedings of the 2022 CHI conference on human factors in computing systems* (pp. 1-19).

Lee, D., Gopal, A., Lee, D., & Shin, D. (2023). Nudging Private Ryan: Mobile microgiving under economic incentives and audience effects. *MIS Quarterly*, 47(3), 1101-1146.

Li, X., & Kim, K. (2024). Impacts of generative AI on user contributions: evidence from a coding Q&A platform. *Marketing Letters*, 1-15.

Li, Z., Wang, G., & Wang, H. J. (2021). Peer effects in competitive environments: Field experiments on information provision and interventions. *MIS Quarterly*, 45(1), 163-191.

Liu, Y. (2006). Word of mouth for movies: Its dynamics and impact on box office revenue. *Journal of Marketing*, 70(3), 74-89.

Mejia, J., Mankad, S., & Gopal, A. (2021). Service quality using text mining: *Measurement and consequences. Manufacturing & Service Operations Management*, 23(6), 1354-1372.

Mimno, D., Wallach, H., Talley, E., Leenders, M., & McCallum, A. (2011, July). Optimizing semantic coherence in topic models. *In Proceedings of the 2011 conference on empirical methods in natural language processing* (pp. 262-272).

Mirsch, T., Lehrer, C., & Jung, R. (2017). Digital nudging: Altering user behavior in digital environments. Proceedings der 13. Internationalen Tagung Wirtschaftsinformatik (WI 2017), 634-648.

Mudambi, S. M., & Schuff, D. (2010). Research note: What makes a helpful online review? A study of customer reviews on Amazon.com. *MIS Quarterly*, 185-200.

Qiao, D., Lee, S. Y., Whinston, A. B., & Wei, Q. (2020). Financial incentives dampen altruism in online prosocial contributions: A study of online reviews. *Information Systems Research*, 31(4), 1361-1375.

Rivera, M., Guo, X., Shan, G., & Qiu, L. (2025). Stimulating Feedback Contributions Using Digital Nudges: A Field Experiment in a Real-Time Mobile Feedback Platform. *MIS Quarterly* 2025

Sachdeva, A., Kim, A., & Dennis, A. R. (2024). Taking the Chat out of Chatbot? Collecting user reviews with Chatbots and web forms. *Journal of Management Information Systems*, 41(1), 146-177.

Saldanha, C. (2023). "Yelp enhances consumer experience with new discovery, review and services features." *Yelp*, https://blog.yelp.com/news/winter-product-release-2024/. Accessed March 26, 2026

Saldanha, C. (2024). "Yelp releases a series of new discovery, contribution, services and AI-powered features." *Yelp*, https://blog.yelp.com/news/winter-product-release-2024/. Accessed March 26, 2026

Schneider, C., Weinmann, M., & Vom Brocke, J. (2018). Digital nudging: guiding online user choices through interface design. *Communications of the ACM*, 61(7), 67-73.

Shankar, R., & Sim, J. (2024). Generative AI on the Loose: Impact of Improved AI and Expert Oversight on Knowledge Sharing. *Available at SSRN.*

Sherwin, K, Dykes, Y. (2025, June 13). "AI Summaries of Reviews." *NN/g*. https://www.nngroup.com/articles/ai-reviews/. Accessed on March 26th, 2026

Shukla, A. D., Goh, J. M., Sun, T., Gao, G., & Agarwal, R. (2025). Effects of Nudging and Privacy Control on Online Physician Reviews: Evidence from a Field Experiment. *Journal of Management Information Systems*, 42(2), 529-563.

Singh, N., Bernal, G., Savchenko, D., & Glassman, E. L. (2023). Where to hide a stolen elephant: Leaps in creative writing with multimodal machine intelligence. *ACM Transactions on Computer-Human Interaction*, 30(5), 1-57.

Su, Y., Wang, Q., Qiu, L., & Chen, R. (2024). Navigating the Sea of Reviews: Unveiling the Effects of Introducing AI-Generated Summaries in E-Commerce. *Available at SSRN.*

Sweller, J. (1988). Cognitive load during problem solving: Effects on learning. Cognitive science, 12(2), 257-285.

Pamuru, V., Kar, W., & Khern-am-nuai, W. (2023). Status Downgrade: The Impact of Losing Status on a User-Generated Content Platform. *Production and Operations Management*, 10591478241279801.

Petridis, S., Diakopoulos, N., Crowston, K., Hansen, M., Henderson, K., Jastrzebski, S., ... & Chilton, L. B. (2023, April). Angle kindling: Supporting journalistic angle ideation with large language models. *In Proceedings of the 2023 CHI conference on human factors in computing systems* (pp. 1-16).

Wang, C., Zhang, X., & Hann, I. H. (2018). Socially nudged: A quasi-experimental study of friends' social influence in online product ratings. *Information Systems Research*, 29(3), 641-655.

Wang, Y., Goes, P., Wei, Z., & Zeng, D. (2019a). Production of online word-of-mouth: Peer effects and the moderation of user characteristics. *Production and Operations Management*, 28(7), 1621-1640.

Wang, Y., Wang, J., & Yao, T. (2019b). What makes a helpful online review? A meta-analysis of review characteristics. *Electronic Commerce Research*, 19(2).

Wang, X., & Zheng, J. (2024) Can Banning AI-generated Content Save User-Generated Q&A Platforms? *Available at SSRN.*

Weinmann, M., Schneider, C., & Brocke, J. V. (2016). Digital nudging. *Business & Information Systems Engineering*, 58(6), 433-436.

Wood, D., Bruner, J. S., & Ross, G. (1976). The role of tutoring in problem solving. *Journal of Child Psychology and Psychiatry*, 17(2), 89-100.

Wiles, E., Munyikwa, Z., & Horton, J. (2025). Algorithmic writing assistance on jobseekers' resumes increases hires. *Management Science*, 71(12), 10144-10164.

Willemsen, L. M., Neijens, P. C., Bronner, F., & De Ridder, J. A. (2011). "Highly recommended!" The content characteristics and perceived usefulness of online consumer reviews. *Journal of Computer-Mediated Communication*, 17(1), 19-38.

Xiao, P., Chen, Y., Bharadwaj, A., & Bao, W. (2022). The effects of information nudges on consumer usage of digital services under three-part tariffs. *Journal of Management Information Systems*, 39(1), 130-158.

Xue, J., Wang, L., Zheng, J., Li, Y., & Tan, Y. (2023). Can ChatGPT Kill User-Generated Q&A Platforms? *Available at SSRN.*

Yang, Y., Zhang, K., & Fan, Y. (2023). sdtm: A supervised Bayesian deep topic model for text analytics. *Information Systems Research*, 34(1), 137-156.

Yin, D., & Bond, S. D. (2021). Anger in consumer reviews: Unhelpful but persuasive? *MIS Quarterly*, 45(3).

Yin, D., Bond, S. D., & Zhang, H. (2014). Anxious or angry? Effects of discrete emotions on the perceived helpfulness of online reviews. *MIS Quarterly*, 38(2), 539-560.

Zeng, Z., Dai, H., Zhang, D. J., Zhang, H., Zhang, R., Xu, Z., & Shen, Z. J. M. (2023). The impact of social nudges on user-generated content for social network platforms. *Management Science*, 69(9), 5189-5208.

Zhao, W., Chen, J. J., Perkins, R., Liu, Z., Ge, W., Ding, Y., & Zou, W. (2015, December). A heuristic approach to determine an appropriate number of topics in topic modeling. *In BMC bioinformatics* (Vol. 16, pp. 1-10).

Zhang, X., & Zhu, F. (2011). Group size and incentives to contribute: A natural experiment at Chinese Wikipedia. *American Economic Review*, 101(4), 1601-1615.

# Guiding without Generating: Artificial Intelligence (AI)-Enabled Topic Nudges in Online Reviews

## APPENDIX

### A. Topic Modeling Using LDA

#### A.1 Determining the optimal number of topics in LDA

To determine the optimal number of topics in LDA, we use various metrics, including the perplexity score (Blei et al., 2003), the rate of perplexity change (RPC) (Zhao et al., 2015), and the coherence score (Mimno et al., 2011). The perplexity score assesses how well a model predicts unseen data, with lower scores indicating a better fit. RPC refines this by evaluating the rate of change in perplexity as the number of topics varies, with lower rates indicating improved fit. The coherence score measures the semantic similarity among top words in a topic, with higher scores signifying more interpretable topics. We vary the number of LDA topics from 2 to 9 and compute these metrics. Based on the comprehensive evaluation reported in Figure 2, we select four topics as optimal.

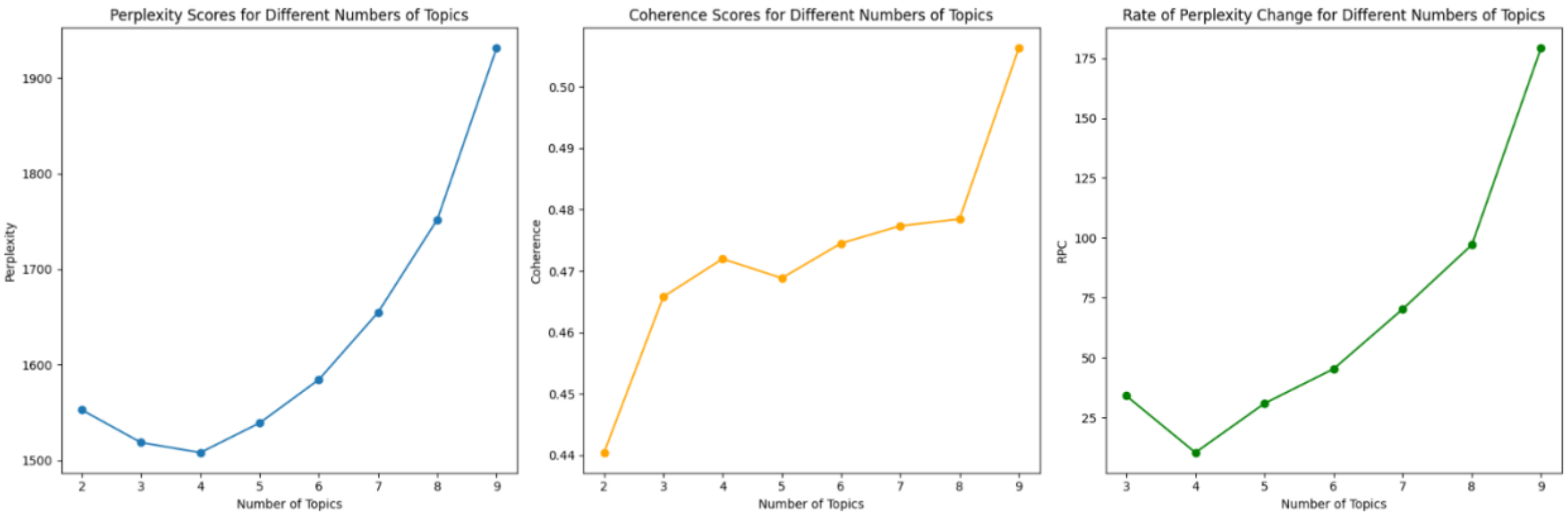

Figure A.1. Evaluation Score for Different Numbers of Topics

**A.2 Top Words and Reviews that Load Most Heavily associated with Each Topic**

**Reviews that load most heavily onto service**

**Keywords: "order", "food", "time", "ask", "wait", "servic", "place", "like", "dont", "tabl", "didnt", "minut", "want", "said", "custom", "come", "came", "peopl", "got", "restaur"**

"The hostess gave me a hard time for having a service animal and accused my service animal of being an ESA. Despite me repeatedly telling them that the animal was not an ESA and did in fact provide a service for my disability, they denied me entry…."

"Very bad customer service at this location - there is a need for training in this area (i mean customer service...Hope this helps truly improve Jimmy Johns customer service experience at this store and in general."

"If I could score this location a 0 stars I would. The workers here have the absolute worst customer service skills I \'ve ever experienced. There is no reason why anybody should have to wait almost 30 minutes to get a Mobile Order…"

**Reviews that load most heavily onto food**

**Key words: "good", "chicken", "order", "fri", "flavor", "like", "food", "dish", "sauc", "tri", "tast", "rice", "place", "got", "meat", "delici", "salad", "roll", "spici", "beef"**

"I've been to Running Goose twice but the first time was almost 4 years ago and unfortunately I don't remember it much…Corn tostada: this was good but I thought it could use a little more balance - maybe a touch of acid and salt or some spice. I did like the charred corn on it though. The tostada shell was fine but I've definitely had better …"

"'Loooove this place and ordered it 3 times in one week…' Some things I tried: ', "* Mapo tofu: one of the BEST mapo tofu I've had. So saucy, savory, and delicious. It was so spicy and it burned my mouth and I cry while I eat it but it's so delicious that I keep ordering it and just crying..."

['This is the first time I ordered from them and definitely not the last time.', "Bbq pork chop is a must have! I don't know if they hooked me up or something but it's such a big bone-in piece of pork chop and it's so juicy and tender! With great sweet Hawaiian bbq flavor…"

**Reviews that load most heavily onto drink**

**Key words: "coffe", "good", "tri", "like", "place", "ice", "cream", "flavor", "sweet", "tea", "love", "got", "sandwich", "breakfast", "cake", "drink", "tast", "order", "park", "delici"**

"Koffeteria is a coffee shop located in the East Downtown. They offer a wide variety of unique drinks and pastries…Overall, I think their classic drinks and pastries are very unique and special…You can definitely tell that the pastries were made well and with lots of creativity."

"A friend of mine recommended we try this place. What intrigued me the most were the free bottles you can get if you check in on Yelp and their souffl√© pancakes. ', 'I bought the Tiramisu souffl√© cake, Ceylon milk tea w/ the tiramisu puff cream + boba + half sugar w less ice, jasmine green milk tea, and the mixed fruit tea…"

"If you aren't familiar with pineapple buns, they are soft buns with a crusty exterior that can be plain or filled with creams or butter…Ice Chilled Milk Tea', '- Ice Chilled Coffee Milk Tea ', '- Lemon Iced Tea'…"

Reviews that load most heavily onto ambiance

**Keywords: "food", "great", "servic", "place", "good", "love", "amaz", "recommend", "delici", "time", "staff", "drink", "friendli", "definit", "best", "nice", "come", "restaur", "experi", "spot"**

"We had a destination wedding 10 years ago and our reception was in the upstairs at Wayfare Tavern. Then a few weeks ago we went back to San Francisco to celebrate our milestone anniversary and we brought our kids to Wayfare for a celebration dinner. ', 'We caught an early bird dinner... The interior is cozy, yet refined, very well-designed and comfortable finishes…"

"This place is AMAZING! Every time I've gone the food and the service has been fantastic…I would highly recommend this place for a date night, the decor is the perfect setting for a romantic lunch or dinner. I would also recommend splitting most of their entrees, the portions are very generous and splitting them ensures you have room for appetizers and/or dessert…"

"I had a fantastic experience visiting Highland Cigar Company! This cigar bar offers an impressive selection of cigars to choose from, making it a perfect spot for cigar aficionados. The atmosphere of the bar is also excellent and creates a relaxing environment to enjoy a good smoke. In addition to the great selection and atmosphere…"

## B. Robustness Checks for Defining Experienced Users

Our results remain robust when classifying users who have earned at least one "Elite" badge since either 2022 or 2015 as experienced users. In Tables B1 and B2, we report the results for users labeled $Expeienced = 1$, when they have earned at least one "Elite User" badge after 2015, and $Expeienced = 0$ otherwise.

Table B1. Robustness Checks for Experienced Users if Elite after 2015

| Dep. Var.: | (1)<br>log(*Service Score*) | (2)<br>log(*Food Score*) | (3)<br>log(*Ambiance Score*) | (4)<br>log(*Service Lengths*) | (5)<br>log(*Food Lengths*) | (6)<br>log(*Ambiance Lengths*) |
|---|---|---|---|---|---|---|
| Panel A: Less Experienced Users | | | | | | |
| Treat*Post | 0.009*** | -0.008*** | -0.001 | 0.074*** | 0.004 | 0.028*** |
| | (0.001) | (0.001) | (0.001) | (0.009) | (0.007) | (0.007) |
| Treat | -0.000 | -0.0003 | -0.000 | -0.006 | -0.004 | |
| | (0.000) | (0.000) | (0.000) | (0.008) | (0.006) | -0.005 (0.006) |
| Num. Obs. | 685,525 | 685,525 | 685,525 | 685,525 | 685,525 | 685,525 |
| Panel B: Experienced Users | | | | | | |
| Treat*Post | -0.001 | -0.003* | 0.004*** | 0.017 | 0.023** | 0.062*** |
| | (0.001) | (0.001) | (0.001) | (0.015) | (0.009) | (0.011) |
| Treat | -0.0004 | 0.0002 | 0.000 | -0.011 | -0.004 | -0.003 |
| | (0.001) | (0.001) | (0.001) | (0.012) | (0.008) | (0.009) |
| Num. Obs. | 290,966 | 290,966 | 290,966 | 290,966 | 290,966 | 290,966 |
| Control Variables, restaurant FE, and calendar week FE included | | | | | | |

*Note. Standard errors in parentheses are robust and clustered at the restaurant level.* * $p < 0.1$, ** $p < 0.05$, *** $p < 0.01$.

Table B2. Robustness Checks for Experienced Users if Elite after 2015

| Dependent Var.: | (1)<br>log($Lengths$) | (2)<br>log($Review\ Volume$) | (3)<br>log($Helpfulness\ Votes$) | (4)<br>log(*Entropy*) |
|---|---|---|---|---|
| Panel A: Less Experienced Users | | | | |
| Treat*Post | 0.046*** | -0.022*** | -0.022*** | 0.001 |
| | (0.004) | (0.002) | (0.003) | (0.001) |
| Treat | -0.004 | -0.005*** | -0.002 | -0.000 |
| | (0.003) | (0.001) | (0.003) | (0.000) |
| Num. Obs. | 685,525 | 685,525 | 685,525 | 685,525 |
| Panel B: Experienced Users | | | | |
| Treat*Post | 0.028*** | -0.008*** | -0.074*** | 0.004*** |
| | (0.005) | (0.002) | (0.007) | (0.001) |
| Treat | -0.005 | -0.002 | -0.001 | -0.000 |
| | (0.005) | (0.002) | (0.006) | (0.001) |
| Num. Obs. | 290,966 | 290,966 | 290,966 | 290,966 |
| Control Variables, restaurant FE, and calendar week FE included | | | | |

*Note: The average length includes only reviews in the English language. Standard errors in parentheses are robust and clustered at the restaurant level.* * $p < 0.1$, ** $p < 0.05$, *** $p < 0.01$.

In Tables B3 and B4, we report the results for users labeled $Expeienced = 1$, when they have earned at least one "Elite User" badge after 2015, and $Expeienced = 0$ otherwise.

Table B3. Robustness Checks for Experienced Users if Elite after 2022

| Dep. Var.: | (1) log(*Service Score*) | (2) log(*Food Score*) | (3) log(*Ambiance Score*) | (4) log(*Service Lengths*) | (5) log(*Food Lengths*) | (6) log(*Ambiance Lengths*) |
|---|---|---|---|---|---|---|
| Panel A: Less Experienced Users | | | | | | |
| Treat*Post | 0.008*** | -0.006*** | -0.002* | 0.079*** | 0.022*** | 0.033*** |
| | (0.001) | (0.001) | (0.001) | (0.009) | (0.007) | (0.007) |
| Treat | -0.000 | -0.0002 | -0.0003 | -0.007 | -0.004 | |
| | (0.001) | (0.000) | (0.000) | (0.008) | (0.006) | -0.004 (0.006) |
| Num. Obs. | 698,269 | 698,269 | 698,269 | 698,269 | 698,269 | 698,269 |
| Panel B: Experienced Users | | | | | | |
| Treat*Post | -0.0004 | -0.003** | 0.004*** | 0.016 | 0.023** | 0.062*** |
| | (0.001) | (0.001) | (0.001) | (0.015) | (0.009) | (0.011) |
| Treat | -0.0004 | 0.0001 | 0.000 | -0.011 | -0.005 | -0.003 |
| | (0.001) | (0.001) | (0.001) | (0.013) | (0.008) | (0.010) |
| Num. Obs. | 271,737 | 271,737 | 271,737 | 271,737 | 271,737 | 271,737 |
| Control Variables, restaurant FE, and calendar week FE included | | | | | | |

*Note. Standard errors in parentheses are robust and clustered at the restaurant level.* * $p < 0.1$, ** $p < 0.05$, *** $p < 0.01$.

Table B4. Robustness Checks for Experienced Users if Elite after 2022

| Dependent Var.: | (1) log($Lengths$) | (2) log($Review\ Volume$) | (3) log($Helpfulness\ Votes$) | (4) log(*Entropy*) |
|---|---|---|---|---|
| Panel A: Less Experienced Users | | | | |
| Treat*Post | 0.055*** | -0.020*** | -0.015*** | 0.002* |
| | (0.004) | (0.002) | (0.003) | (0.001) |
| Treat | -0.004 | -0.004*** | -0.001 | -0.000 |
| | (0.003) | (0.001) | (0.003) | (0.000) |
| Num. Obs. | 698,269 | 698,269 | 698,269 | 698,269 |
| Panel B: Experienced Users | | | | |
| Treat*Post | 0.028*** | -0.012*** | -0.068*** | 0.004*** |
| | (0.006) | (0.002) | (0.007) | (0.001) |
| Treat | -0.004 | -0.002 | -0.001 | -0.0002 |
| | (0.005) | (0.002) | (0.006) | (0.001) |
| Num. Obs. | 271,737 | 271,737 | 271,737 | 271,737 |
| Control Variables, restaurant FE, and calendar week FE included | | | | |

*Note: The average length includes only reviews in the English language. Standard errors in parentheses are robust and clustered at the restaurant level.* * $p < 0.1$, ** $p < 0.05$, *** $p < 0.01$.

## C. Effects of the AI-Enabled Topic Tudges with a Relative Time Model

In Table C1, we report the estimated results using a relative time model (Equation (6) in the paper).

Table C1. Effects of the AI-Enabled Topic Nudges with a Relative Time Model

| Dependent Var.: | (1) log($Length$) | (2) log($Review Volume$) | (3) log(*Service Score*) | (4) log(*Food Score*) | (5) log(*Ambiance Score*) | (6) log(*Service Lengths*) | (7) log(*Food Lengths*) | (8) log(*Ambiance Lengths*) | (9) log($Helpfulness Vote$) | (10) log(*Entropy*) |
|---|---|---|---|---|---|---|---|---|---|---|
| Pre (≥8) | -0.012 | 0.014*** | -0.0007 | 0.003 | -0.002 | -0.005 | 0.006 | -0.024* | 0.025*** | -0.0002 |
| | (0.008) | (0.003) | (0.002) | (0.002) | (0.002) | (0.018) | (0.014) | (0.014) | (0.007) | (0.002) |
| Pre (-7) | -0.009 | 0.009** | -0.0009 | -0.001 | 0.003 | 0.003 | -0.009 | 0.009 | 0.009 | 0.002 |
| | (0.010) | (0.004) | (0.002) | (0.002) | (0.003) | (0.024) | (0.019) | (0.018) | (0.009) | (0.002) |
| Pre (-6) | -0.009 | 0.005 | -0.001 | -0.0002 | 0.001 | -0.014 | -0.007 | -0.007 | 0.006 | -0.0003 |
| | (0.010) | (0.004) | (0.002) | (0.002) | (0.003) | (0.024) | (0.018) | (0.018) | (0.009) | (0.002) |
| Pre (-5) | -0.023** | 0.003 | -0.005* | 0.001 | 0.003 | -0.054* | -0.014 | 0.003 | 0.012 | -0.0006 |
| | (0.010) | (0.004) | (0.002) | (0.002) | (0.003) | (0.024) | (0.018) | (0.018) | (0.009) | (0.002) |
| Pre (-4) | -0.019* | 0.008** | 6.66e-6 | -0.003 | 0.003 | -0.018 | -0.032* | 0.008 | 0.022** | -0.0007 |
| | (0.010) | (0.004) | (0.002) | (0.002) | (0.003) | (0.024) | (0.019) | (0.018) | (0.009) | (0.002) |
| Pre (-3) | -0.017* | 0.005 | -0.004 | -0.0002 | 0.004 | -0.041* | -0.019 | 0.013 | 0.011 | -0.0008 |
| | (0.010) | (0.004) | (0.002) | (0.002) | (0.003) | (0.024) | (0.018) | (0.018) | (0.009) | (0.002) |
| Pre (-2) | 0.007 | 0.004 | 0.0002 | -0.0007 | 0.0006 | 0.014 | 0.005 | 0.008 | 0.007 | 0.000 |
| | (0.010) | (0.004) | (0.002) | (0.002) | (0.003) | (0.024) | (0.019) | (0.018) | (0.009) | (0.002) |
| Pre (-1) | Baseline Period Omitted | | | | | | | | | |
| Post (0) | 0.014 | -0.006 | 0.0008 | 0.002 | -0.003 | 0.007 | 0.024 | -0.017 | -0.004 | 0.0003 |
| | (0.010) | (0.004) | (0.002) | (0.002) | (0.003) | (0.024) | (0.019) | (0.018) | (0.009) | (0.002) |
| Post (1) | 0.007 | -0.001 | -0.005** | 0.006*** | -0.0004 | -0.040* | 0.042** | 0.011 | -0.005 | 0.005** |
| | (0.010) | (0.004) | (0.002) | (0.002) | (0.003) | (0.024) | (0.018) | (0.018) | (0.009) | (0.002) |
| Post (2) | 0.020** | -0.002 | 0.002 | -0.002 | -0.0008 | 0.008 | 0.010 | 0.023 | -0.009 | -0.002 |
| | (0.010) | (0.004) | (0.003) | (0.002) | (0.003) | (0.024) | (0.018) | (0.018) | (0.009) | (0.002) |
| Post (3) | 0.029*** | -0.006 | 0.002 | -0.002 | 0.0004 | 0.034 | 0.024 | 0.036** | -0.027*** | 0.004* |
| | (0.010) | (0.004) | (0.002) | (0.002) | (0.003) | (0.024) | (0.019) | (0.018) | (0.008) | (0.002) |
| Post (4) | 0.028*** | -0.011*** | 0.004 | -0.006** | 0.002 | 0.039 | -0.003 | 0.042** | -0.024*** | 0.002 |
| | (0.010) | (0.004) | (0.002) | (0.002) | (0.003) | (0.024) | (0.018) | (0.018) | (0.009) | (0.002) |
| Post (5) | 0.022** | -0.015*** | 0.005** | -0.004* | -0.001 | 0.039* | 0.002 | 0.003 | -0.032*** | 0.0002 |
| | (0.010) | (0.004) | (0.002) | (0.002) | (0.003) | (0.024) | (0.018) | (0.018) | (0.009) | (0.002) |
| Post (6) | 0.043*** | -0.031*** | 0.006** | -0.003 | -0.003 | 0.062*** | 0.033* | 0.016 | -0.008 | 0.004* |
| | (0.010) | (0.004) | (0.003) | (0.003) | (0.003) | (0.024) | (0.019) | (0.018) | (0.009) | (0.002) |
| Post (7) | 0.042*** | -0.013*** | 0.001 | -0.003 | 0.003 | 0.033 | 0.029 | 0.065*** | -0.027*** | 0.005** |
| | (0.010) | (0.004) | (0.002) | (0.002) | (0.003) | (0.024) | (0.018) | (0.018) | (0.006) | (0.002) |
| Post (8) | 0.036*** | -0.022*** | 0.005*** | -0.005** | -0.0007 | 0.043** | 0.010 | 0.022. | -0.004 | 0.001 |
| | (0.007) | (0.003) | (0.002) | (0.002) | (0.002) | (0.018) | (0.013) | (0.013) | (0.009) | (0.002) |

| Treat | Yes | Yes | Yes | Yes | Yes | Yes | Yes | Yes | Yes | Yes |
|---|---|---|---|---|---|---|---|---|---|---|
| Control | Yes | Yes | Yes | Yes | Yes | Yes | Yes | Yes | Yes | Yes |
| Restaurant FE | Yes | Yes | Yes | Yes | Yes | Yes | Yes | Yes | Yes | Yes |
| Calendar Week FE | Yes | Yes | Yes | Yes | Yes | Yes | Yes | Yes | Yes | Yes |
| Num. Obs. | 847,902 | 847,902 | 847,902 | 847,902 | 847,902 | 847,902 | 847,902 | 847,902 | 847,902 | 847,902 |
| $R^2$ | 0.040 | 0.037 | 0.041 | 0.036 | 0.037 | 0.035 | 0.035 | 0.039 | 0.032 | 0.037 |

*Note: Standard errors in parentheses are robust and clustered at the restaurant level.* $^{*}$ $p < 0.1$, $^{**}$ $p < 0.05$, $^{***}$ $p < 0.01$.

## D. Alternative Control Groups

Table G1. Beauty Category as a Control Group

| Dep. Var.: | log($Review\ Volume$) | log($Helpfulness\ Votes$) | log($Lengths$) |
|---|---|---|---|
| Year2023 | -0.015*** | -0.103*** | -0.045*** |
| | (0.004) | (0.007) | (0.008) |
| Treat*Post | 0.017*** | 0.001 | -0.021*** |
| | (0.003) | (0.006) | (0.007) |
| Treat*Year2023 | 0.029*** | 0.009 | -0.018** |
| | (0.004) | (0.008) | (0.009) |
| Post*Year2023 | 0.001 | -0.023*** | 0.034*** |
| | (0.004) | (0.008) | (0.010) |
| Treat*Post*Year2023 | -0.027*** | -0.022** | 0.011 |
| | (0.004) | (0.009) | (0.010) |
| Restaurant FE | Yes | Yes | Yes |
| Calendar Week FE | Yes | Yes | Yes |
| Num. Obs. | 1,050,625 | 1,050,625 | 1,050,625 |
| $R^2$ | 0.457 | 0.217 | 0.140 |

*Note: Standard errors in parentheses are robust and clustered at the restaurant level. * p < 0.1, ** p < 0.05, *** p < 0.01.*

Table G1. Google Reviews as a Control Group

| | (1) | (2) |
|---|---|---|
| Dependent Var.: | log($Review\ Volume$) | log($Helpfulness\ Votes$) |
| Treat*Post | -0.012*** | -0.061*** |
| | (0.001) | (0.003) |
| Platform-Restaurant FE | Yes | Yes |
| Week FE | Yes | Yes |
| Num. Obs. | 1,349,235 | 1,349,235 |
| $R^2$ | 0.511 | 0.243 |

*Note: Standard errors in parentheses are robust and clustered at the restaurant level. * p < 0.1, ** p < 0.05, *** p < 0.01.*

## E. Additional Analysis for Updated Review Helpfulness Votes

Table E1. Additional Analysis for Updated Review Helpfulness Votes

| | (1) |
|---|---|
| Dependent Var.: | log($Helpfulness\ Votes$) |
| Treat*Post | -0.023*** |
| | (0.004) |
| Treat | -0.090*** |
| | (0.003) |
| Control | Yes |
| Restaurant FE | Yes |
| Calendar Week FE | Yes |
| Num. Obs. | 781,126 |
| $R^2$ | 0.198 |

*Note: Standard errors in parentheses are robust and clustered at the restaurant level.* * $p < 0.1$, ** $p < 0.05$, *** $p < 0.01$.

## F. Additional Analysis at Both the Review and Review Writer Level

Table F1. Additional Analysis at the Review Level

| | Review-level | | |
|---|---|---|---|
| Dep. Var.: | log($Length$) | log($Helpfulness\ Votes$) | log($Entropy$) |
| Treat*Post | 0.050*** | -0.014*** | 0.003*** |
| | (0.003) | (0.003) | (0.001) |
| Treat | -0.070*** | -0.080*** | 0.002** |
| | (0.003) | (0.002) | (0.001) |
| Restaurant FE | Yes | Yes | Yes |
| Calendar Week FE | Yes | Yes | Yes |
| Num. Obs. | 1,559,528 | 1,559,528 | 1,559,528 |
| $R^2$ | 0.088 | 0.104 | 0.103 |

*Note. FE: fixed effect; Robust standard errors are in parentheses; *p < 0.1; **p < 0.05; ***p < 0.01.*

Table F2. Additional Analysis at the Review Level

| | Review-level | | | | | |
|---|---|---|---|---|---|---|
| | (1) | (2) | (3) | (4) | (5) | (6) |
| Dep. Var.: | log(*Service Score*) | log(*Food Score*) | log(*Ambiance Score*) | log(*Service Lengths*) | log(*Food Lengths*) | log(*Ambiance Lengths*) |
| Treat*Post | 0.005*** | -0.005*** | 0.002*** | 0.077*** | 0.024*** | 0.060*** |
| | (0.001) | (0.001) | (0.001) | (0.007) | (0.006) | (0.005) |
| Treat | -0.010*** | -0.012*** | 0.022*** | -0.122*** | -0.176*** | 0.097*** |
| | (0.001) | (0.001) | (0.001) | (0.006) | (0.007) | (0.004) |
| Restaurant FE | Yes | Yes | Yes | Yes | Yes | Yes |
| Calendar Week FE | Yes | Yes | Yes | Yes | Yes | Yes |
| Num. Obs. | 1,559,528 | 1,559,528 | 1,559,528 | 1,559,528 | 1,559,528 | 1,559,528 |
| $R^2$ | 0.225 | 0.280 | 0.260 | 0.134 | 0.190 | 0.275 |

*Note. FE: fixed effect; Robust standard errors are in parentheses; *p < 0.1; **p < 0.05; ***p < 0.01.*

Table F3. Additional Analysis at the Review Writer Level

| | Review Writer-level | | | |
|---|---|---|---|---|
| | (1) | (2) | (3) | (4) |
| Dep. Var.: | log(*Length*) | log(*Review Volume*) | log(*Helpfulness Votes*) | log(*Entropy*) |
| Treat*Post | 0.030*** | -0.004*** | -0.050*** | 0.002** |
| | (0.004) | (0.002) | (0.003) | (0.0010) |
| Treat | -0.011*** | -0.018*** | -0.059*** | 0.005*** |
| | (0.003) | (0.001) | (0.003) | (0.0008) |
| Review Writer FE | Yes | Yes | Yes | Yes |
| Calendar Week FE | Yes | Yes | Yes | Yes |
| Num. Obs. | 1,302,066 | 1,302,066 | 1,302,066 | 1,559,528 |
| $R^2$ | 0.829 | 0.494 | 0.750 | 0.703 |

*Note: Standard errors in parentheses are robust and clustered at the review writer level.* $*p < 0.1$; $**p < 0.05$; $***p < 0.01$.

Table F4. Additional Analysis at the Review Writer Level

| | Review Writer-level | | | | | |
|---|---|---|---|---|---|---|
| | (1) | (2) | (3) | (4) | (5) | (6) |
| Dep. Var.: | log(*Service Score*) | log(*Food Score*) | log(*Ambiance Score*) | log(*Service Lengths*) | log(*Food Lengths*) | log(*Ambiance Lengths*) |
| Treat*Post | -0.0003 | -0.003** | 0.004*** | 0.023** | 0.024*** | 0.047*** |
| | (0.001) | (0.001) | (0.001) | (0.010) | (0.007) | (0.009) |
| Treat | -0.004*** | -0.011*** | 0.016*** | -0.037*** | -0.056*** | 0.115*** |
| | (0.0009) | (0.0009) | (0.0009) | (0.008) | (0.006) | (0.008) |
| Review Writer FE | Yes | Yes | Yes | Yes | Yes | Yes |
| Calendar Week FE | Yes | Yes | Yes | Yes | Yes | Yes |
| Num. Obs. | 1,302,066 | 1,302,066 | 1,302,066 | 1,302,066 | 1,302,066 | 1,302,066 |
| $R^2$ | 0.782 | 0.765 | 0.777 | 0.744 | 0.814 | 0.708 |

*Note. Standard errors in parentheses are robust and clustered at the review writer level.* $*p < 0.1$; $**p < 0.05$; $***p < 0.01$.

## G. Alternative Specifications for Equation (1) in the Manuscript

Table G1. Difference-in-Differences on Non-Detrended Outcomes

| Dep. Var.: | (1)<br>log($Lengths$) | (2)<br>log($Review\ Volume$) | (3)<br>log($Helpfulness\ Votes$) | (3)<br>log($Entropy$) |
|---|---|---|---|---|
| Treat*Post | 0.048*** | -0.025*** | -0.037*** | 0.002*** |
| | (0.004) | (0.002) | (0.004) | (0.007) |
| Treat | -0.069*** | 0.011*** | -0.104*** | -0.002*** |
| | (0.003) | (0.001) | (0.003) | (0.0006) |
| Control | Yes | Yes | Yes | Yes |
| Restaurant FE | Yes | Yes | Yes | Yes |
| Calendar Week FE | Yes | Yes | Yes | Yes |
| Num. Obs. | 851,019 | 851,019 | 851,019 | 851,019 |
| $R^2$ | 0.132 | 0.460 | 0.208 | 0.146 |

*Note: The average length includes only reviews in the English language. Standard errors in parentheses are robust and clustered at the restaurant level.* $^{*}$ *p < 0.1,* $^{**}$ *p < 0.05,* $^{***}$ *p < 0.01.*

Table G2. Difference-in-Differences on Non-Detrended Outcomes

| Dep. Var.: | (1)<br>log(*Service Score*) | (2)<br>log(*Food Score*) | (3)<br>log(*Ambiance Score*) | (4)<br>log(*Service Lengths*) | (5)<br>log(*Food Lengths*) | (6)<br>log(*Ambiance Lengths*) |
|---|---|---|---|---|---|---|
| Treat*Post | 0.006*** | -0.005*** | -0.001 | 0.056*** | 0.020*** | 0.033*** |
| | (0.001) | (0.001) | (0.001) | (0.008) | (0.006) | (0.006) |
| Treat | -0.010*** | -0.012*** | 0.024*** | -0.110*** | -0.106*** | 0.103*** |
| | (0.001) | (0.001) | (0.001) | (0.007) | (0.005) | (0.005) |
| Control | Yes | Yes | Yes | Yes | Yes | Yes |
| Restaurant FE | Yes | Yes | Yes | Yes | Yes | Yes |
| Calendar Week FE | Yes | Yes | Yes | Yes | Yes | Yes |
| Num. Obs. | 851,019 | 851,019 | 851,019 | 851,019 | 851,019 | 851,019 |
| $R^2$ | 0.294 | 0.337 | 0.315 | 0.177 | 0.241 | 0.340 |

*Note. Standard errors in parentheses are robust and clustered at the restaurant level.* $^{*}$ *p < 0.1,* $^{**}$ *p < 0.05,* $^{***}$ *p < 0.01.*

Table G3. Controlling for City-Specific Time Trends

| Dep. Var.: | (1)<br>log(*Lengths*) | (2)<br>log(*Review Volume*) | (3)<br>log(*Helpfulness Votes*) | (3)<br>log(*Entropy*) |
|---|---|---|---|---|
| Treat*Post | 0.048*** | -0.025*** | -0.037*** | 0.002*** |
| | (0.004) | (0.002) | (0.004) | (0.007) |
| Treat | -0.006* | 0.007*** | -0.004 | -0.000 |
| | (0.003) | (0.001) | (0.003) | (0.000) |
| Control | Yes | Yes | Yes | Yes |
| City-Specific Time Trends | Yes | Yes | Yes | Yes |
| Restaurant FE | Yes | Yes | Yes | Yes |
| Calendar Week FE | Yes | Yes | Yes | Yes |
| Num. Obs. | 847,902 | 847,902 | 847,902 | 847,902 |
| $R^2$ | 0.039 | 0.037 | 0.032 | 0.037 |

*Note: The average length includes only reviews in the English language. Standard errors in parentheses are robust and clustered at the restaurant level.* $^{*}$ *p < 0.1,* $^{**}$ *p < 0.05,* $^{***}$ *p < 0.01.*

Table G4. Controlling for City-Specific Time Trends

| Dep. Var.: | (1)<br>log(*Service Score*) | (2)<br>log(*Food Score*) | (3)<br>log(*Ambiance Score*) | (4)<br>log(*Service Lengths*) | (5)<br>log(*Food Lengths*) | (6)<br>log(*Ambiance Lengths*) |
|---|---|---|---|---|---|---|
| Treat*Post | 0.006*** | -0.005*** | -0.001 | 0.055*** | 0.020*** | 0.033*** |
| | (0.001) | (0.001) | (0.001) | (0.008) | (0.006) | (0.006) |
| Treat | -0.000 | -0.0001 | -0.0002 | -0.010 | -0.007 | -0.005 |
| | (0.001) | (0.001) | (0.001) | (0.007) | (0.005) | (0.005) |
| Control | Yes | Yes | Yes | Yes | Yes | Yes |
| City-Specific Time Trends | Yes | Yes | Yes | Yes | Yes | Yes |
| Restaurant FE | Yes | Yes | Yes | Yes | Yes | Yes |
| Calendar Week FE | Yes | Yes | Yes | Yes | Yes | Yes |
| Num. Obs. | 847,902 | 847,902 | 847,902 | 847,902 | 847,902 | 847,902 |
| $R^2$ | 0.041 | 0.035 | 0.037 | 0.035 | 0.035 | 0.038 |

*Note. Standard errors in parentheses are robust and clustered at the restaurant level.* $^{*}$ *p < 0.1,* $^{**}$ *p < 0.05,* $^{***}$ *p < 0.01.*